\documentclass[preprint]{aastex}

\pdfoutput=1

\usepackage{verbatim}
\usepackage{color}

\slugcomment{Published in ApJ 756, 80 (2012)}

\shorttitle{HCN, CH$_3$OH, and Rotational Temperature in Comet
Hartley 2} \shortauthors{Drahus et al.}

\begin{document}

\title{The Sources of HCN and CH$_3$OH and the Rotational Temperature in Comet \mbox{103P/Hartley 2} from Time-Resolved Millimeter Spectroscopy\altaffilmark{*}}

\author{Micha{\l} Drahus, David Jewitt, and Aur\'elie Guilbert-Lepoutre}
\affil{Department of Earth and Space Sciences, University of
California at Los Angeles,\\ Los Angeles, USA}
\email{mdrahus@ucla.edu}

\author{Wac{\l}aw Waniak}
\affil{Astronomical Observatory, Jagiellonian University, Krak\'ow,
Poland}

\and

\author{Albrecht Sievers}
\affil{Instituto de Radio Astronom\`ia Milim\'etrica, Granada,
Spain}

\altaffiltext{*}{Based on observations carried out with the
IRAM~30~m telescope. IRAM is supported by INSU/CNRS (France), MPG
(Germany), and IGN (Spain).}

\begin{abstract}
One of the least understood properties of comets is the
compositional structure of their nuclei, which can either be
homogeneous or heterogeneous. The nucleus structure can be
conveniently studied at millimeter wavelengths, using
velocity-resolved spectral time series of the emission lines,
obtained simultaneously for multiple molecules as the body rotates.
Using this technique, we investigated the sources of CH$_3$OH and
HCN in comet \mbox{103P/Hartley 2}, the target of NASA's
\emph{EPOXI} mission, which had an exceptionally favorable
apparition in late 2010. Our monitoring with the IRAM~30~m telescope
shows short-term variability of the spectral lines caused by nucleus
rotation. The varying production rates generate changes in
brightness by a factor of~4 for HCN and by a factor of~2 for
CH$_3$OH, and they are remarkably well correlated in time. With the
addition of the velocity information from the line profiles, we
identify the main sources of outgassing: two jets, oppositely
directed in a radial sense, and icy grains, injected into the coma
primarily through one of the jets. The mixing ratio of CH$_3$OH and
HCN is dramatically different in the two jets, which evidently shows
large-scale chemical heterogeneity of the nucleus. We propose a
network of identities linking the two jets with morphological
features reported elsewhere, and postulate that the chemical
heterogeneity may result from thermal evolution. The model-dependent
average production rates are $3.5\times10^{26}$ molec~s$^{-1}$ for
CH$_3$OH and $1.25\times10^{25}$ molec~s$^{-1}$ for HCN, and their
ratio of~28 is rather high but not abnormal. The rotational
temperature from CH$_3$OH varied strongly, presumably due to nucleus
rotation, with the average value being 47~K.
\end{abstract}


\section{Introduction}\label{Sec-Intro}

Comets are icy remnants holding clues about the formation and
evolution of the Solar System. Depending on the region of formation
in the protosolar nebula, they are currently stored in two main
reservoirs: the Oort cloud and the Kuiper belt. The Oort cloud is a
source of long-period comets and (probably) Halley-type comets
\citep{Lev96}. It has been suggested that Oort cloud comets formed
in the giant planet region and were subsequently ejected to the
periphery of the Solar System \citep[e.g.][]{Don04}, but also that
some may have been captured from other stellar systems while the Sun
was in its birth cluster \citep{Lev10}. The Kuiper belt is a source
of Jupiter-family comets \citep[e.g.][]{Dun04}. Kuiper belt comets
presumably formed just beyond the orbit of Neptune where they
continue to orbit. By studying comets from different reservoirs we
can probe the different environments in which they formed and also
better understand their role in the Solar System as the suppliers of
water and organics.

\mbox{103P/Hartley 2} (hereafter 103P) is a Jupiter-family comet
which currently has a \mbox{6.47-year} orbital period and perihelion
at 1.06~AU. On UT~2010 Oct.~20.7 it reached the minimum geocentric
distance of only 0.12~AU, making by far the closest approach to the
Earth since its discovery \citep{Har86}, and becoming visible to the
naked eye. Shortly after, on UT~2010 Nov.~4.5832, the comet was
visited by NASA's \emph{EPOXI} spacecraft, which provided detailed
images and spectra of the nucleus and its closest surroundings
\citep{AHe11a}. Both the Earth-based data, taken at an unusually
favorable geometry, and the unique observations carried out by the
spacecraft, create an exceptional platform for new groundbreaking
investigations.

One of the most fundamental problems of cometary science is the
compositional structure of the nucleus, which holds unique
information about the formation and evolution of comets. A nucleus
that condensed in one place would be, at least initially,
homogeneous and compositionally similar to that region of the
protosolar nebula. In contrast, a heterogeneous composition would
suggest formation from smaller ``cometesimals'' which accumulated
into comets in the early Solar System. Because of the expected
radial migration of cometesimals \citep{Wei77}, they could originate
at different heliocentric distances in the protosolar disk and hence
have different chemical compositions. The above interpretation can
be biased, to some extent, for thermally evolved comets, in which
depletion in the most volatile ices may occur non-uniformly
\citep{Gui11}.

Both homogeneous \citep[e.g.][]{Del07} and heterogeneous
\citep[e.g.][]{Gib07} compositions have been suggested for different
comets based on ground-based IR spectroscopy of the emission lines.
The best information, however, have come from the spatially resolved
molecular images of water (H$_2$O) and carbon dioxide (CO$_2$),
obtained for comets \mbox{9P/Tempel 1} and 103P within the
\emph{Deep Impact} and \emph{EPOXI} missions, respectively
\citep{Fea07,AHe11a}. These observations showed that each comet
emits the two molecules from distinct sources at different locations
on their nuclei. But ground-based IR spectroscopy of 103P
\citep{Mum11,Del11} did not provide any compelling evidence for
similar differences among other molecules, including methanol
(CH$_3$OH) and hydrogen cyanide (HCN). We address this issue in
detail in this work, using our own millimeter-wavelength
observations of CH$_3$OH and HCN in this comet.

Millimeter-wavelength spectroscopy is a powerful tool with which to
investigate comets because it is sensitive to parent molecules
through their rotational transitions and because the spectra are
velocity-resolved. A time series of velocity-resolved spectra,
obtained simultaneously for multiple molecules, reveal their
production rates and line-of-sight kinematics over the course of
nucleus rotation. We can thereby identify whether these molecules
originate from the same source(s) or from different sources
(compositional homogeneity vs. heterogeneity), and in this way gain
rare and valuable insights into the compositional structure of the
nucleus. Moreover, millimeter spectroscopy provides excellent
diagnostics of the rotational temperature in the coma, which can be
derived from simultaneous observations of different transitions from
the same molecule.

In the present paper, we continue the exploration of our mm/submm
spectroscopic observations of 103P obtained in late 2010. Earlier we
quantified the rotation state of the nucleus based on an extensive
monitoring of the HCN line variability observed at multiple
telescopes \citep[][hereafter Paper~I]{Dra11}. This time we focus on
a small but unique subset of data from a single instrument to
investigate the sources of CH$_3$OH and HCN. Using the former
molecule, we also constrain time-resolved rotational temperature.
Our findings presented in Paper~I have been accounted for in the
present work. In particular, we now average the spectra in longer
blocks (typically 1~hr vs. 15~min in Paper~I), to increase the
signal-to-noise ratio. This change is motivated by the fact that the
nucleus rotation period, equal to 18.33~hr at the epoch of
observations \citep[cf.~Paper~I; also e.g.][]{AHe11a}, is long
enough to prevent any significant variability on timescales shorter
than one hour. Moreover, we analyze the spectra in 0.15~km~s$^{-1}$
bins (vs. 0.10 and 0.25~km~s$^{-1}$ in Paper~I), which is limited by
the native resolution available for CH$_3$OH. While we are generally
consistent with the methodology used in Paper~I, in the current work
we calculate the line parameters from a narrower window (from
$-1.75$ to $+1.75$~km~s$^{-1}$ instead of from $-2$ to
$+2$~km~s$^{-1}$) to further improve the signal-to-noise ratio, and
we derive the molecular production rates using a revised rotational
temperature (47~K from CH$_3$OH at the epoch of observations,
instead of 30~K obtained previously from the long-term monitoring of
HCN at various telescopes).

\section{Observations and Data Reduction}\label{Sec-Obs}

We took observations at the 30~m millimeter telescope on Pico Veleta
(Spain), operated by the Institut de Radioastronomie Millim\'etrique
(IRAM). On three consecutive nights: UT~2010 Nov.~3.03--3.35,
Nov.~4.03--4.34, and Nov.~5.02--5.35, we obtained time series of
velocity-resolved spectra of CH$_3$OH and HCN. The middle moment of
these two time series (hereafter the epoch of observations) is
UT~2010 Nov.~4.1908, close to the moment of the \emph{EPOXI}
encounter that occurred 9.4~hr later. Our observations cover the
epochs immediately before (37.2--29.5~hr and 13.4--5.8~hr) and after
(10.4--18.3~hr) the flyby, while the moment of the encounter could
not be covered because of the geographic longitude of the telescope.
The weather was consistently good and stable, with the median zenith
opacity at 225~GHz equal to~0.18, and we encountered basically no
technical problems. At the epoch of observations the helio- and
geocentric distances were 1.0631 and 0.1546~AU, respectively, and
the geocentric phase angle was 58.82$\degr$. We consider these
parameters to be valid for the entire time series given that the
changes in geometry were very small. HCN data were also obtained one
night earlier, on UT~2010 Nov.~2.06--2.37 (60.7--53.1~hr before the
flyby), but we excluded them from the analysis because no
counterpart spectra of CH$_3$OH were taken at that time;
nonetheless, this additional HCN dataset is presented along with the
main data and we refer to these spectra on two occasions.

The two molecules were observed simultaneously with the Eight Mixer
Receiver (EMIR): HCN~\emph{J}(3--2) in the E3 band at 265.886434~GHz
and five lines of CH$_3$OH in the E1 band centered at 157.225~GHz.
EMIR is a state-of-the-art sideband-separating dual-polarization
instrument having a typical receiver temperature of 85~K in E3 and
50~K in E1. Spectral decomposition was performed simultaneously by
the Versatile Spectrometer Array (VESPA) and the Wideband Line
Multiple Autocorrelator (WILMA). For the analysis we chose the
highest-resolution data from VESPA. The sections connected to E3
have \mbox{39.1-kHz} spectral-channel spacing (resolution $R =
7\times10^6$) and \mbox{36.0-MHz} bandwidth (921 spectral channels
per polarization) and the sections connected to E1 provide
\mbox{78.1-kHz} spacing (resolution $R = 2\times10^6$) and
\mbox{71.6-MHz} bandwidth (917 spectral channels per polarization).
In each band the two polarization channels were aligned to better
than 2$\arcsec$ on the sky, which we concluded from frequent
pointing calibrations on compact continuum sources (see further).
Table~\ref{Tab1} summarizes the transition and telescopic constants
relevant to this work.

\placetable{Tab1}

All observations were taken in \emph{position-switching} mode in
which the whole antenna moves between the source position (ON) and a
sky reference position (OFF). The offset between the two was
15$\arcmin$ in azimuth, which secures OFF to be free (for all
practical purposes) of cometary contribution \citep{Dra10}, and is
still sufficiently close to ON to serve as a good reference giving
relatively flat baselines. The integration times at ON and OFF were
equal to 15~sec, which was chosen based on established instrumental
and atmospheric characteristic timescales. We consistently took 8
\emph{subscans} (i.e. \mbox{ON--OFF} pairs) per spectrum, giving the
total integration time at ON equal to 2~min, the total ON+OFF
integration time of 4~min, and the effective observation time of
$\sim5$~min per spectrum (including the overhead for antenna
operations). The chopper-wheel calibration \citep{Uli76,Kra97} was
performed every 3 spectra ($\sim15$~min); it was used by the system
to automatically scale the signal in terms of the antenna
temperature, which we further converted to the main-beam brightness
temperature $T_\mathrm{mB}$ using the main-beam efficiency
(Table~\ref{Tab1}). Cometary observations were most often taken in
$\sim$\mbox{1-hr} blocks. For the purpose of the present work, we
averaged the spectra within the blocks and altogether from the two
polarization channels to improve the signal-to-noise ratio. We used
statistical weights inversely proportional to the square of the
system temperature, which is a good proxy of noise when the
integration time and spectral channels are the same for all input
spectra. We obtained the total of 21 such spectra for each molecule,
7 per night, supplemented by 7 spectra of HCN from the first night.

As the moments of observation we use the middle times of the blocks
as measured by the telescope clock (i.e. not corrected for the
travel time of light). With each moment of observation we associate
the nucleus rotation phase, calculated with the constant rotation
period $P=18.33$~hr obtained for the epoch of observations from our
\emph{dynamical solution} presented in Paper~I. We also use the
\emph{three-cycle} nomenclature from Paper~I. The rotation phases
are calculated with respect to the moment of the \emph{EPOXI}
encounter, UT~2010 Nov.~4.5832 as measured by the spacecraft clock
(i.e. at the comet), at which we harbor the middle phase of the
\emph{three-cycle} system, i.e. phase 0.5 of \emph{Cycle~B} (also
consistent with Paper~I). We decided not to use the full
\emph{dynamical solution} for simplicity, to ensure a strictly
linear relation between the phases and the moments of observation,
taking advantage of the fact that both the changes in the rotation
period and the changes in the travel time of light are negligibly
small in the considered time interval. Consequently, even though the
current system of rotation phases is somewhat simplified compared to
Paper~I, for all practical purposes it gives consistent values (the
maximum phase difference between the systems is $\sim$0.002 at the
beginning of the main dataset and $\sim$0.005 at the beginning of
the supplementary dataset from the first night, which are comparable
to the phase errors resulting from the uncertainty of the
\emph{dynamical solution}).

The blocks were preceded by measurements of pointing (and less
frequently focus) corrections on nearby compact continuum sources
and snapshot spectral observations on nearby molecular line
standards. The RMS pointing consistency was typically at the level
of 2$\arcsec$ in each axis, while the gain fluctuations rarely
exceeded 10\%. The control system of the 30~m telescope calculates
the positions of comets in real time, assuming Keplerian orbits. We
used the osculating elements obtained for the dates of observation
with the JPL \emph{Horizons}
system\footnote{\url{http://ssd.jpl.nasa.gov/?horizons}}
\citep{Gio97}. A relative radial-velocity scale was obtained for
each line from the absolute frequency scale through the classical
Doppler law: zero velocity corresponds to the transition rest
frequency, negative velocities to higher frequencies (blueshift),
and positive velocities to lower frequencies (redshift). Topocentric
Doppler corrections were applied automatically in real time.

\begin{figure}
\epsscale{1.00} \plotone{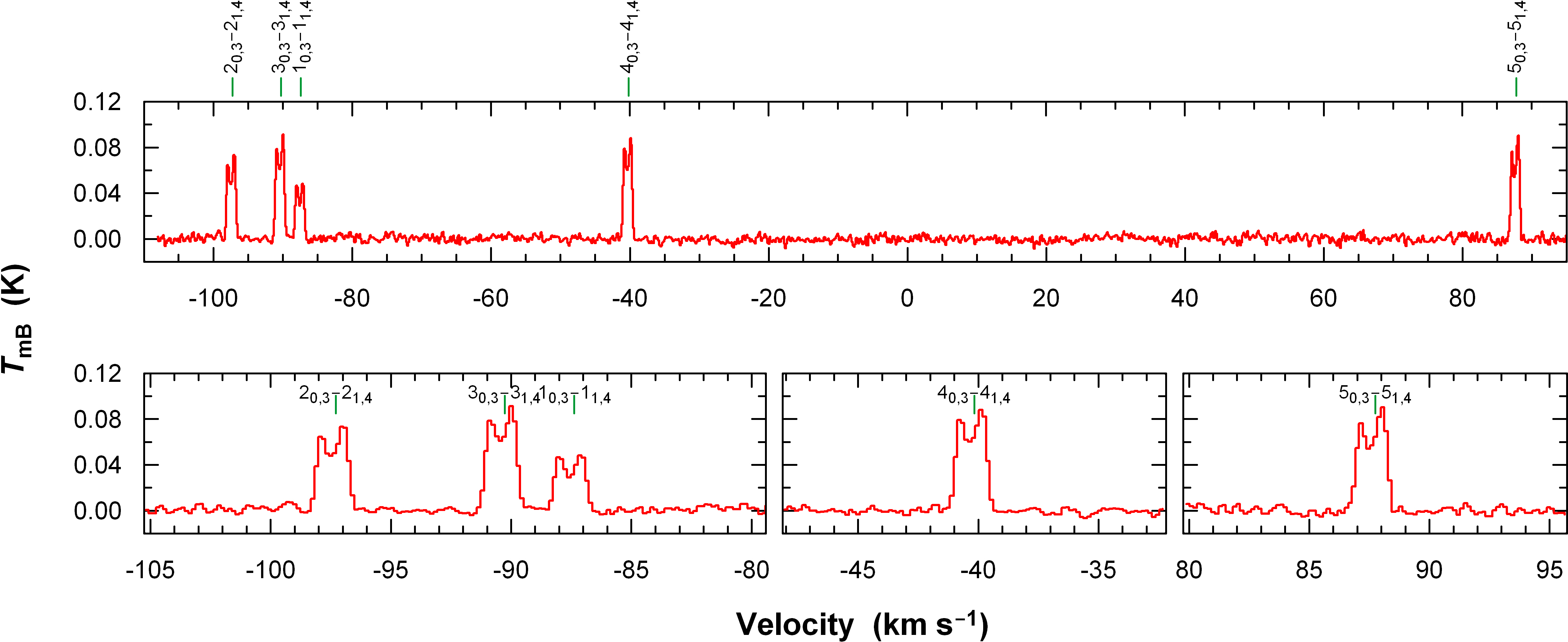} \caption{Mean
spectrum of CH$_3$OH resulting from all 21 spectra in our time
series. The zero velocity corresponds to the rest frequency of
157.225~GHz. The transitions are labeled and the rest velocities are
indicated by the short vertical lines. The \emph{top panel} shows
the full spectrum whereas the \emph{bottom panels} show close-up
views of the line profiles.\label{Fig-AvgMethanol}}
\end{figure}

The spectral baseline was calculated for each spectrum and
separately for each of the observed CH$_3$OH lines. We used a linear
least-squares fit in the interval between $-10$ and $-3$~km~s$^{-1}$
and between $+3$ and $+10$~km~s$^{-1}$, except for the close group
of three CH$_3$OH lines (Fig.~\ref{Fig-AvgMethanol}) for which a
common baseline was calculated; in this case the baseline intervals
were taken with respect to the outer lines. We used only full bins
with native widths inside these intervals, however, bad channels,
with signal exceeding a $3\sigma$ limit, were iteratively rejected.
Then the signal scatter about the baseline was used to calculate
noise RMS in the channels and the baseline was subtracted
(cf.~Paper~I). Finally, each spectrum was rebinned to the standard
velocity-resolution of 0.15~km~s$^{-1}$ ($R = 2\times10^6$), very
close to the native resolution of CH$_3$OH, and the signal error was
propagated to the new channels.

\begin{figure}
\epsscale{0.5} \plotone{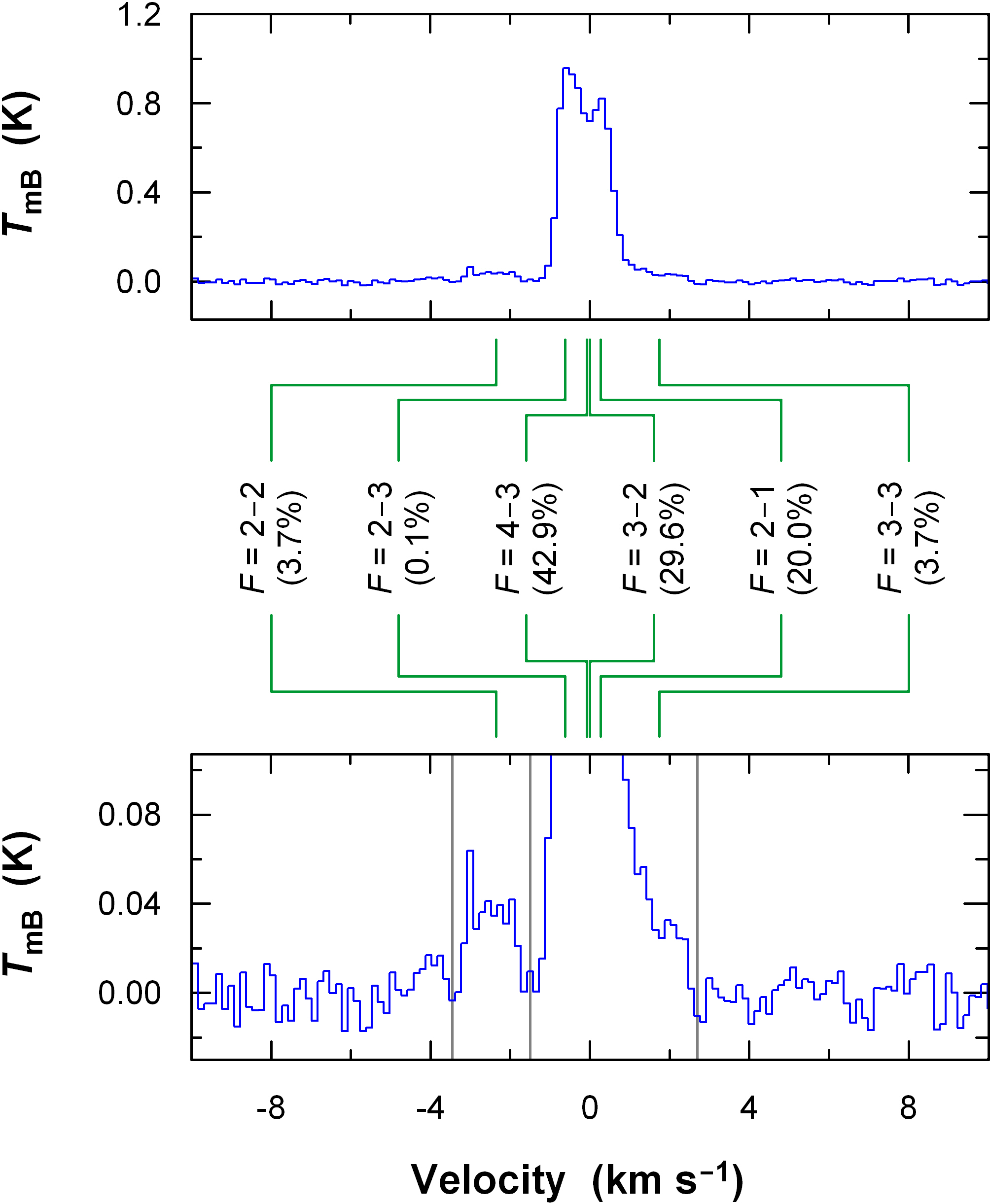} \caption{Mean
spectrum of \mbox{HCN \emph{J}(3--2)} resulting from the 21 spectra
in our time series. The \emph{top panel} shows the complete line
profile whereas the \emph{bottom panel} shows a close-up view of the
baseline. The solid lines between the panels indicate the velocities
of the hyperfine components taken from the Cologne Database for
Molecular Spectroscopy \citep[][]{Mul05}; they are labeled and their
theoretical branching ratios are given in the brackets. It is
evident that, in addition to the three strongest hyperfine
components which build up the line, two faint components are also
visible: one fully resolved at $-2.3545$~km~s$^{-1}$ ($F=2-2$) and
the other one in the red wing of the line at $+1.7394$~km~s$^{-1}$
($F=3-3$). The vertical lines in the \emph{bottom panel} indicate
the velocity ranges in which we measured the component ratio
(Section~\ref{Sec-Sources-ProdRates}).\label{Fig-AvgHCN}}
\end{figure}

We also additionally averaged the spectra in two different ways to
further improve the signal-to-noise ratio: (\emph{i}) for CH$_3$OH
we calculated a time series of mean line profiles, obtained from all
five lines, and (\emph{ii}) for both molecules we calculated their
mean profiles (Figs.~\ref{Fig-AvgMethanol} and~\ref{Fig-AvgHCN})
upon averaging the spectra in the two time series. We used weights
inversely proportional to the square of the signal error in the
spectral channels. The former approach provides no information about
line-to-line behavior but minimizes the noise in the time series and
enables us to better analyze temporal variations. The latter gives
no information about temporal behavior but minimizes the noise
across the bandwidth and is naturally useful to derive ``average''
characteristics of the comet. Note that the result of approach
(\emph{i}) may be difficult to interpret when different lines of the
same molecule have different shapes. This can happen when, for
example, the predominant formation regions are different for the
observed lines and the gas kinematics varies strongly across the
coma (Drahus et~al., in prep.). In our data, however, all the five
lines display practically the same average shapes
(Fig.~\ref{Fig-AvgMethanol}), and hence we assume that they are also
the same in each individual spectrum (this is difficult to verify
because of the much higher noise in the individual spectra). In this
way, using the result of approach (\emph{ii}) we validated approach
(\emph{i}).

\placefigure{Fig-AvgMethanol}

\placefigure{Fig-AvgHCN}

The line profiles are parameterized by their area
\mbox{$\int\!\!T_\mathrm{mB}\mathrm{d}v$} and median velocity $v_0$,
which we derived from the interval between $-1.75$ and
$+1.75$~km~s$^{-1}$. Their errors were estimated from 500
simulations following our Monte Carlo approach, which we used to
propagate the signal noise and also the uncertainty from imperfect
pointing whenever relevant (see Paper~I for details). As the errors
we took the RMS deviations from the measured values, calculated for
the positive and negative sides separately whenever the difference
was significant.

We interpret the observations with the aid of three basic physical
quantities characterizing cometary gas: rotational temperature
$T_\mathrm{rot}$, production rate $Q$, and median radial (i.e.
line-of-sight) component of flow velocity $v_\mathrm{rad}$. These
quantities were derived from the line parameters using the simple
model described in Appendix~\ref{Ap-Model} and fed with the
constants from Table~\ref{Tab1}. It is important to realize that the
absolute values of these quantities are uncertain due to several
simplifying assumptions in this approach. Nevertheless, while we
provide these absolute values, we focus on their temporal
variations, which are affected to a much lesser extent \citep[cf.
the discussion in][where essentially the same approach was
used]{Dra10}. The errors were consistently derived from the
variation of these quantities in the simulated spectra, as outlined
above, except for the errors on $T_\mathrm{rot}$ (further discussed
in Section~\ref{Sec-Trot}). Note that such errors do not include any
other sources of uncertainty, resulting from e.g. data quantization
and calibration, or from the model assumptions and parameters. In
the next sections we analyze these quantities and also the complete
line profiles.

\section{Rotational Temperature}\label{Sec-Trot}

\begin{figure}
\epsscale{1.0} \plotone{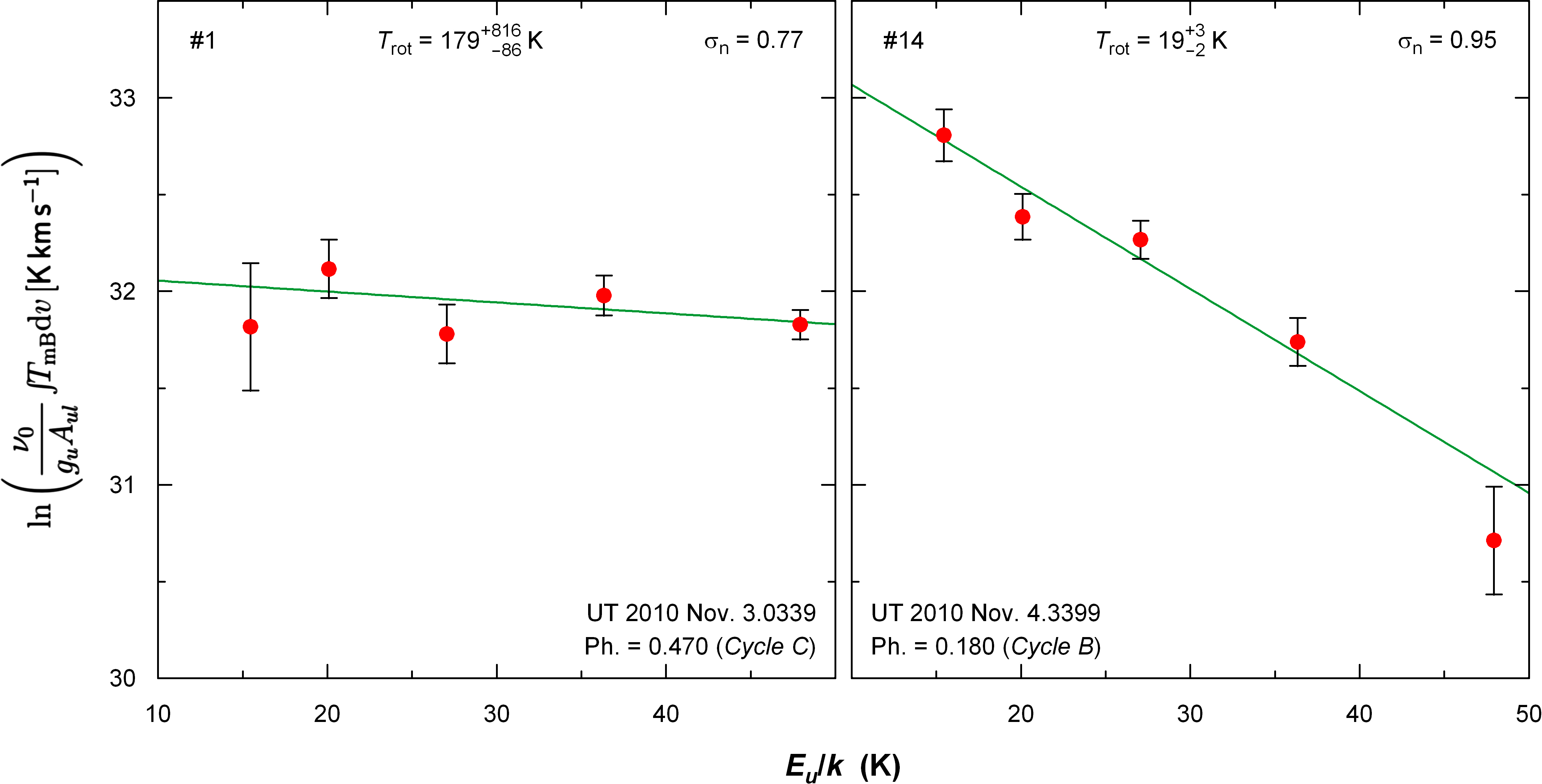} \caption{Examples
of individual rotational diagrams derived from the time series of
CH$_3$OH. The selected diagrams correspond to the maximum
(\emph{left panel}) and minimum (\emph{right panel}) temperatures
$T_\mathrm{rot}$, calculated from the slopes of the weighted linear
least-squares fits (solid lines). Note that the error bars
associated with the individual data points in both diagrams do not
include the uncertainty of the telescope pointing, which affects all
five lines almost identically in the framework of the assumed model
scenario, and therefore does not affect the slope (although it does
affect the zero level). We also show a normalized standard deviation
from the fit $\sigma_\mathrm{n}$ (for an ideal fit
$\sigma_\mathrm{n} = 0$, for deviations ideally consistent with the
errors $\sigma_\mathrm{n} = 1$, and when the measurements deviate
from the model $\sigma_\mathrm{n} > 1$). The complete set of
rotational diagrams is presented in Fig.~\ref{Fig-AllRotDiag}
(Appendix~\ref{Ap-Data}).\label{Fig-ExRotDiag}}
\end{figure}

We applied the \emph{rotational diagram} technique \citep[see
Appendix~\ref{Ap-Model}; also e.g.][]{Boc94} to our time series of
CH$_3$OH to determine the rotational temperature $T_\mathrm{rot}$
and its temporal variation. Selected examples of the rotational
diagrams are presented in Fig.~\ref{Fig-ExRotDiag}. The errors on
$T_\mathrm{rot}$ were not derived directly from the temperature
variation in the simulated spectra (cf. Section~\ref{Sec-Obs}), but
indirectly, from the variation of the rotational-diagram slopes in
these simulations. This change is motivated by the fact that some of
the simulations generated for the flattest rotational diagrams (i.e.
implying the highest $T_\mathrm{rot}$) yield marginally positive
slopes (i.e. non-physical temperatures), however, the temperature
limits calculated from the slope RMS are physical in all cases. Note
that the errors on $T_\mathrm{rot}$ do not account for deviations
from the model (predicting linear rotational diagrams) but are
entirely established by the noise in our data (cf.
Section~\ref{Sec-Obs}).

\placefigure{Fig-ExRotDiag}

According to \citet{Biv02a}, the observed group of transitions at
$\sim157$~GHz yields a rotational temperature very close to the
kinetic temperature of the inner coma. The two temperatures are
strictly equal in \emph{Local Thermodynamic Equilibrium} (LTE), in
which the energy levels are populated according to the Boltzmann
distribution, and the resulting rotational diagrams are linear.
Indeed, the majority of our rotational diagrams are
indistinguishable, within the error bars, from being linear,
although in some cases we observe strong nonlinearities or
accidental deviations. While we cannot exclude instrumental effects
(e.g. such as imperfect removal of the baselines) in the spectra
behind these problematic diagrams, the possibility of large
deviations from the Boltzmann energy-level distributions at these
particular rotation phases cannot be ruled out at this stage.

\begin{figure}
\epsscale{0.5} \plotone{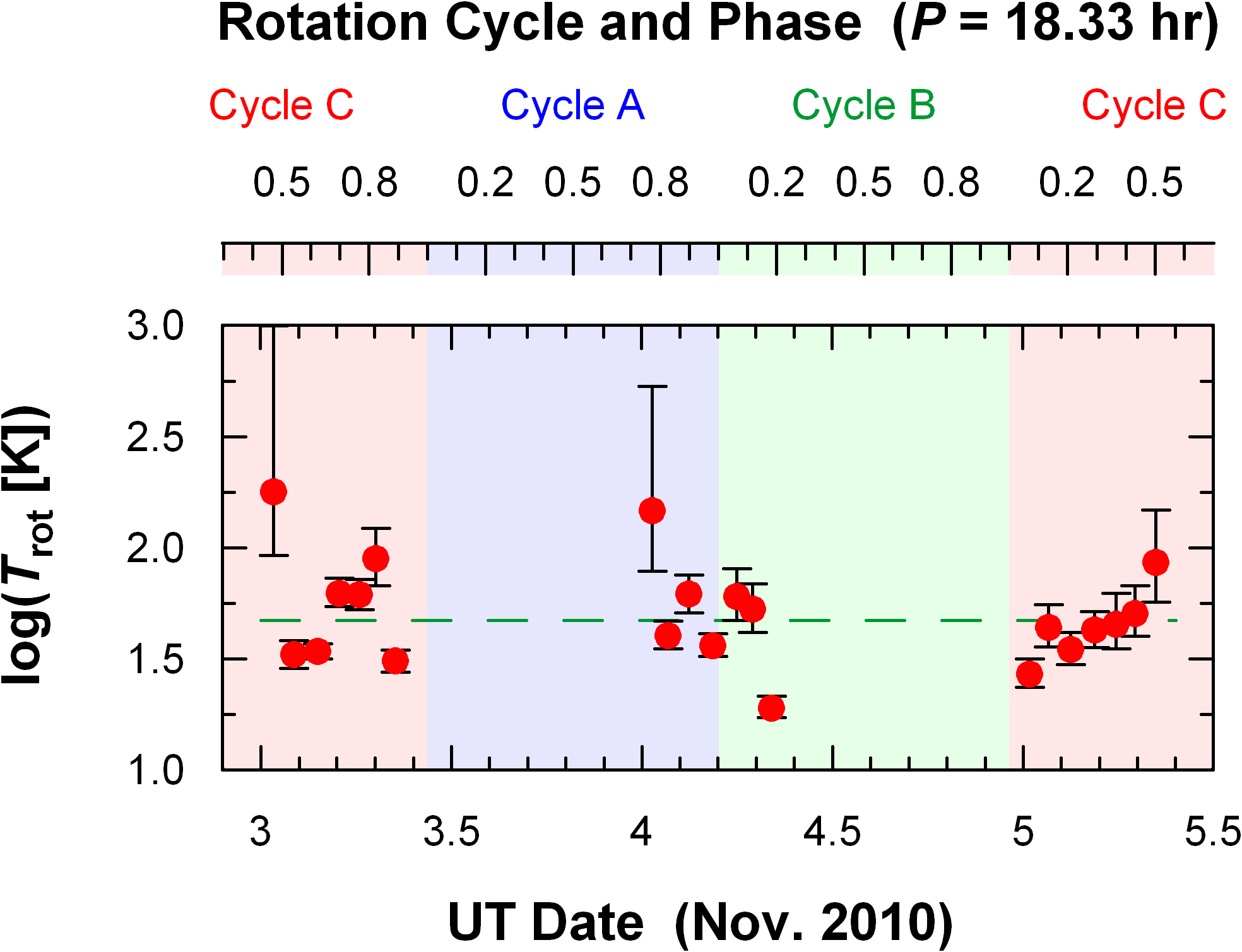} \caption{Variation
of the logarithmic rotational temperature $\log(T_\mathrm{rot})$
with time, derived from the time series of CH$_3$OH. The dashed line
indicates the temperature level of 47~K obtained from the mean
spectrum (Figs.~\ref{Fig-AvgMethanol} and~\ref{Fig-AvgRotDiag}). The
additional top axis shows the nucleus rotation phase and
\emph{three-cycle} component (cf.~Paper~I), and the latter is also
coded by the background color. The \emph{EPOXI} flyby occurred on
UT~2010 Nov.~4.5832, corresponding to phase 0.5 of
\emph{Cycle~B}.\label{Fig-EvolTrot}}
\end{figure}

\begin{figure}
\epsscale{0.5} \plotone{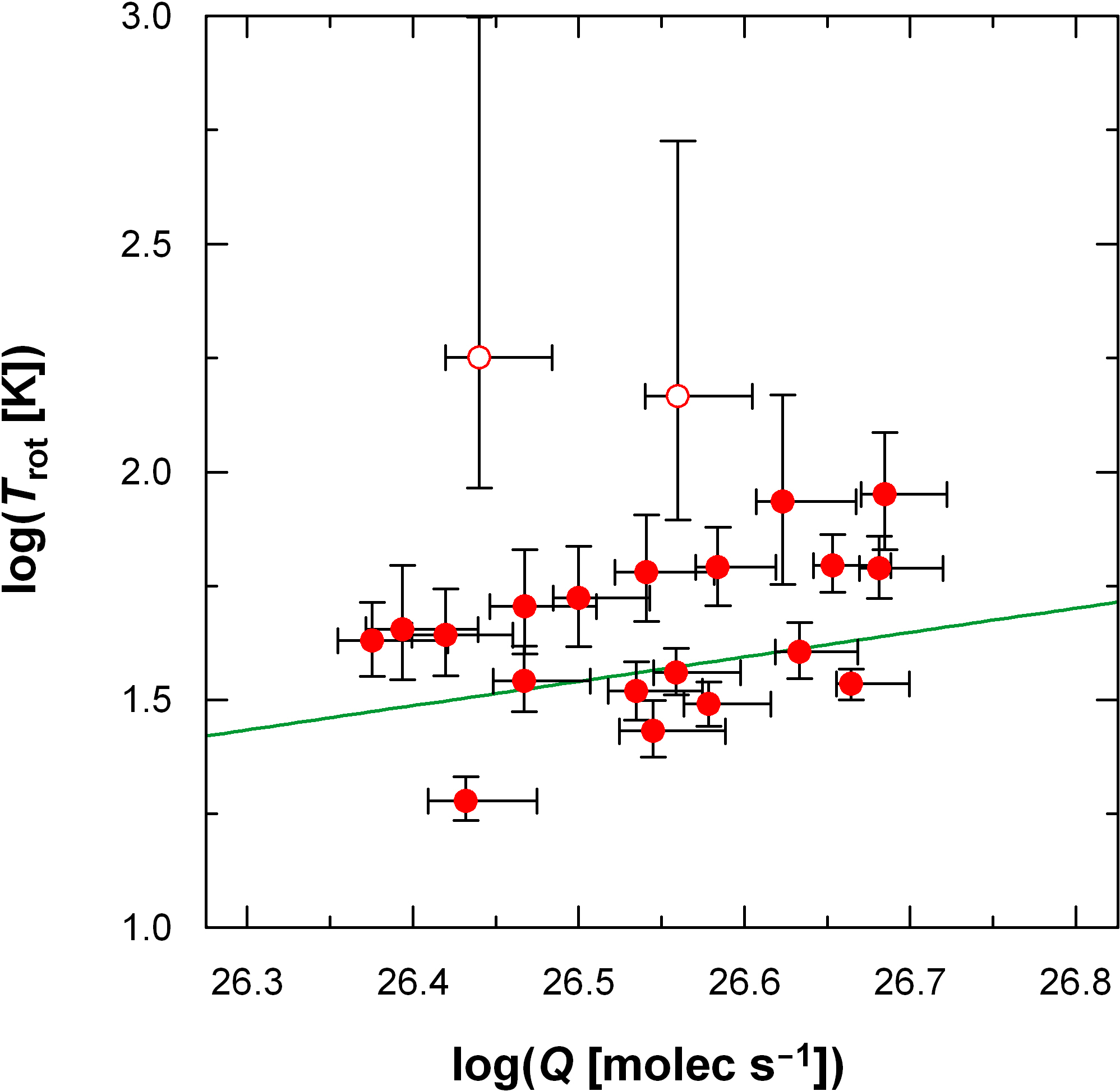} \caption{Tentative
positive correlation of the rotational temperature~$T_\mathrm{rot}$
(Fig.~\ref{Fig-EvolTrot}) and the production rate~$Q$
(Fig.~\ref{Fig-EvolProdRates}), derived from the time series of
CH$_3$OH, and presented in log--log scale. The weighted linear
least-squares fit (solid line) is added to guide the eye. It was
obtained with the weights iteratively calculated from relevant
(positive or negative) sides of the vertical and horizontal error
bars. The fit has a slope of $0.54_{-0.24}^{+0.19}$, where the
errors were estimated from the variation of the slope in the set of
simulated spectra (cf.~Section~\ref{Sec-Obs}). The two
highest-temperature data points (open symbols) were rejected from
the fit because many of the simulations yield non-physical
temperatures for these points and for the same reason five
simulations (from the set of 500) were omitted from the error
estimation, as they yield non-physical temperatures for other data
points. We conclude that this positive correlation must be real
given that negative slopes appear only in 1.6\% of the simulations.
We also note that the two rejected data points have a small
influence on the slope, which is equal to $0.50$ for the complete
dataset.\label{Fig-TrotVsQ}}
\end{figure}

The temporal behavior of $T_\mathrm{rot}$ is presented in
Fig.~\ref{Fig-EvolTrot}. We see large-amplitude variations that seem
to generally follow the rotation-modulated production rate (see
further Fig.~\ref{Fig-EvolProdRates} and the discussion in
Section~\ref{Sec-Sources-ProdRates}), although the correlation is
not strict. While, again, occasional instrumental effects might have
affected this trend, it seems extremely unlikely that the entire
variability is an artifact. Instead, we believe that the variations
are physical and their correlation with the production rate is real
(Fig.~\ref{Fig-TrotVsQ}), but additional fluctuations in the
rotational temperature are superimposed on the regular trend. It is
interesting to note that although a positive correlation of these
two quantities has been predicted by theory \citep[e.g.][]{Com04},
only the correlations in long-term trends, primarily controlled by
changing heliocentric distance, have been reported for individual
objects to date \citep[e.g.][]{Biv02a}. Our result suggests that the
rotational temperature can be positively correlated with the
production rate in a single object also when the received solar
energy flux is constant.

\placefigure{Fig-EvolTrot}

\placefigure{Fig-TrotVsQ}

\begin{figure}
\epsscale{0.5} \plotone{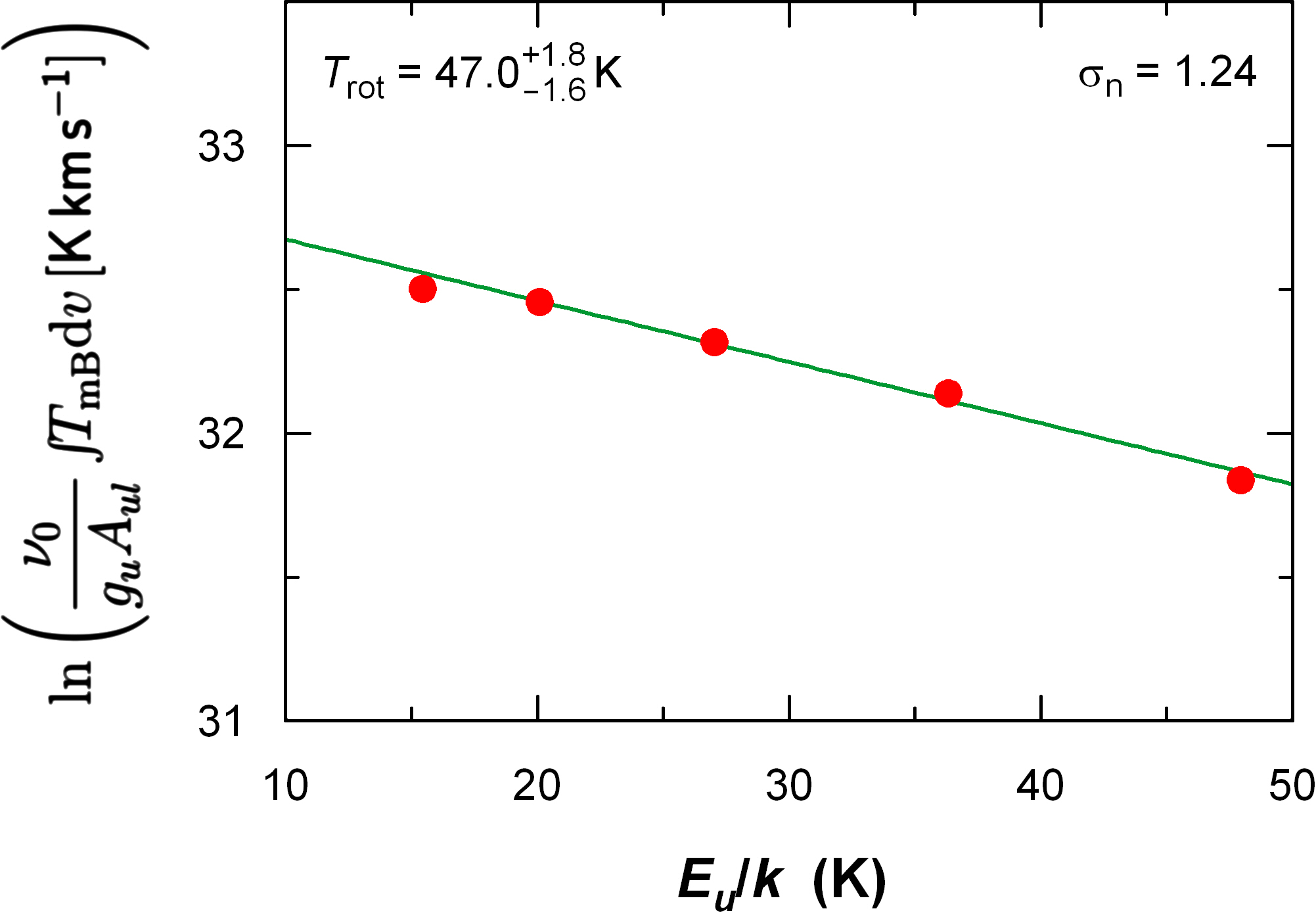}
\caption{Rotational diagram for the mean spectrum of CH$_3$OH from
Fig.~\ref{Fig-AvgMethanol}. The error bars are smaller than the
symbol size. Other details are the same as in
Fig.~\ref{Fig-ExRotDiag}.\label{Fig-AvgRotDiag}}
\end{figure}

We also calculated a rotational diagram for the mean spectrum from
Fig.~\ref{Fig-AvgMethanol}. The diagram, presented in
Fig.~\ref{Fig-AvgRotDiag}, is consistent with the temperature of
$47.0^{+1.8}_{-1.6}$~K. While this value is some 30~K lower than the
rotational temperatures from IR spectroscopy \citep{Mum11,Del11},
the latter were derived from a much smaller volume surrounding the
nucleus, where the gas has presumably the highest temperature
\citep[e.g.][]{Com04}, and hence this difference should be expected
and in fact was often noticed in the past \citep[e.g.][]{Dra10}. It
is important to realize that the gas observed by our beam was
presumably highly non-isothermal, both in time (see above) and
across the coma \citep[e.g.][]{Com04}, even if it locally satisfied
LTE. Moreover, we cannot exclude that non-thermal processes
significantly contributed to the overall excitation scheme,
especially in the outer part of the observed coma. Therefore,
interpretation of the rotational temperature in such an environment
is highly problematic and some of these problems will be discussed
by Drahus et~al. (in prep.).

\placefigure{Fig-AvgRotDiag}

\section{Sources of HCN and CH$_3$OH}\label{Sec-Sources}

\subsection{Average Production Rates and Temporal Variability}\label{Sec-Sources-ProdRates}

\begin{figure}
\epsscale{0.5} \plotone{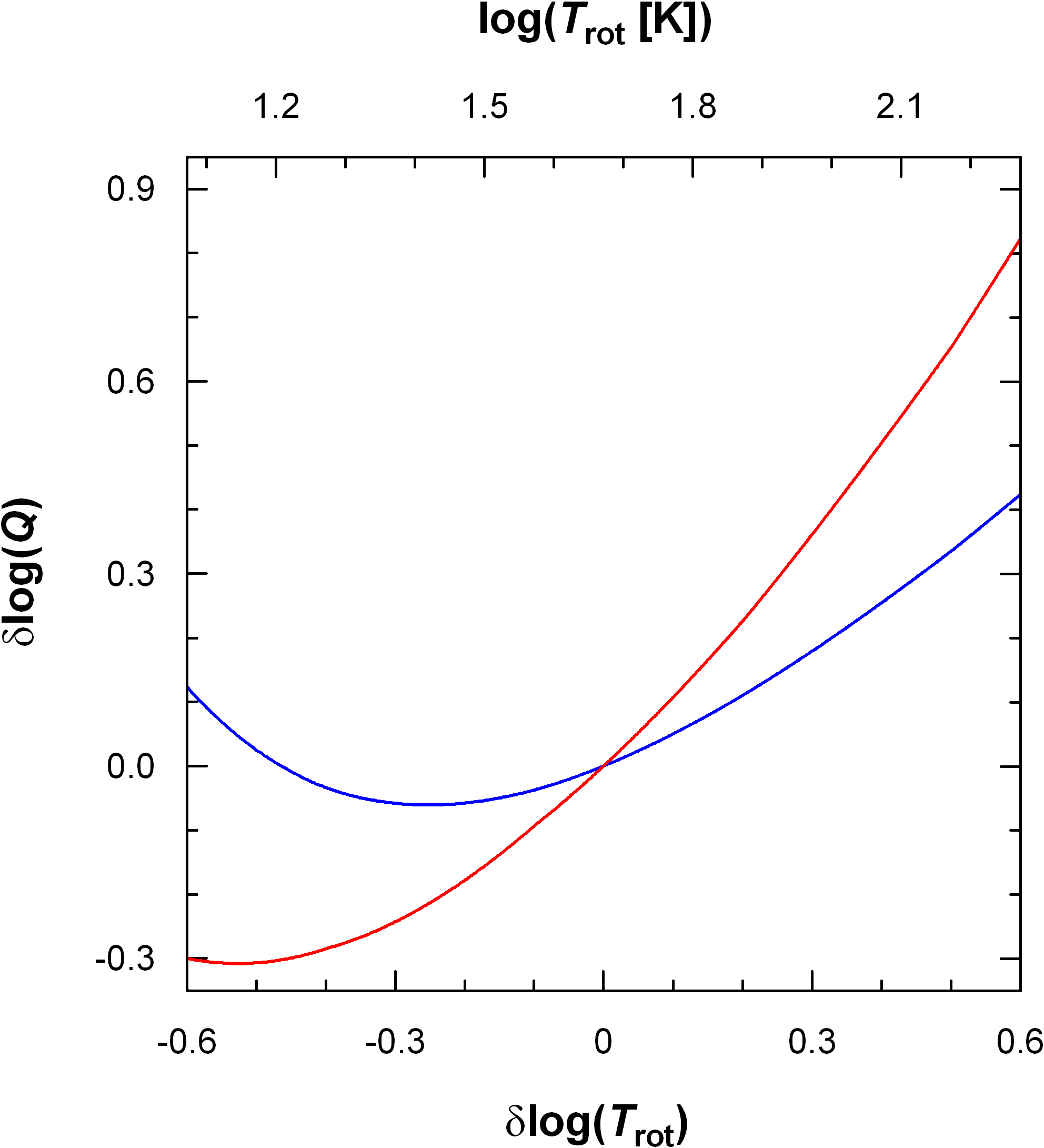}
\caption{Dependence of the derived production rate $Q$ on the
adopted rotational temperature $T_\mathrm{rot}$, presented as
logarithmic offsets applicable to CH$_3$OH (red) and HCN (blue). The
absolute logarithmic temperature scale is given by the upper
horizontal axis.\label{Fig-ModelQVsTrot}}
\end{figure}

Both the average and the instantaneous production rates $Q$ were
derived from the line areas
\mbox{$\int\!\!T_\mathrm{mB}\mathrm{d}v$} assuming a constant gas
expansion velocity $v_\mathrm{gas} = 0.8$~km~s$^{-1}$ and using the
average rotational temperature $T_\mathrm{rot} = 47$~K
(Section~\ref{Sec-Trot}). At this point we refrain from using the
instantaneous $T_\mathrm{rot}$ for the calculation of the
instantaneous $Q$, because of the relatively large errors on the
individual temperatures (especially on the highest ones), the
uncertain reason(s) for the occasional nonlinearities of the
rotational diagrams, and some uncertainty as to the cause(s) of the
observed variability (cf.~Section~\ref{Sec-Trot}). In
Fig.~\ref{Fig-ModelQVsTrot} we show how the derived production rates
depend on the adopted rotational temperature. We also note that in
the framework of our simple model the production rate is a linear
function of the line area, and that we used constant in time
conversion factors for the two molecules
(cf.~Appendix~\ref{Ap-Model}).

\placefigure{Fig-ModelQVsTrot}

The derived average absolute production rates are $3.5\times10^{26}$
molec~s$^{-1}$ for CH$_3$OH and $1.25\times10^{25}$ molec~s$^{-1}$
for HCN. The production-rate ratio is 28, which is some 50\% higher
than the most typical values found in comets from millimeter
spectroscopy \citep[][]{Biv02b}. The derived ratio is also
noticeably higher than the values of 6--9 resulting from the
measurements of 103P by \citet{Mum11} and \citet{Del11} obtained in
the infrared \citep[and updated with the revised IR production rates
of CH$_3$OH by][]{Vil12}, but such an inconsistency should again be
no surprise, given all the differences in data acquisition and
modeling, and also the natural variability of this comet.

\begin{figure}
\epsscale{0.5} \plotone{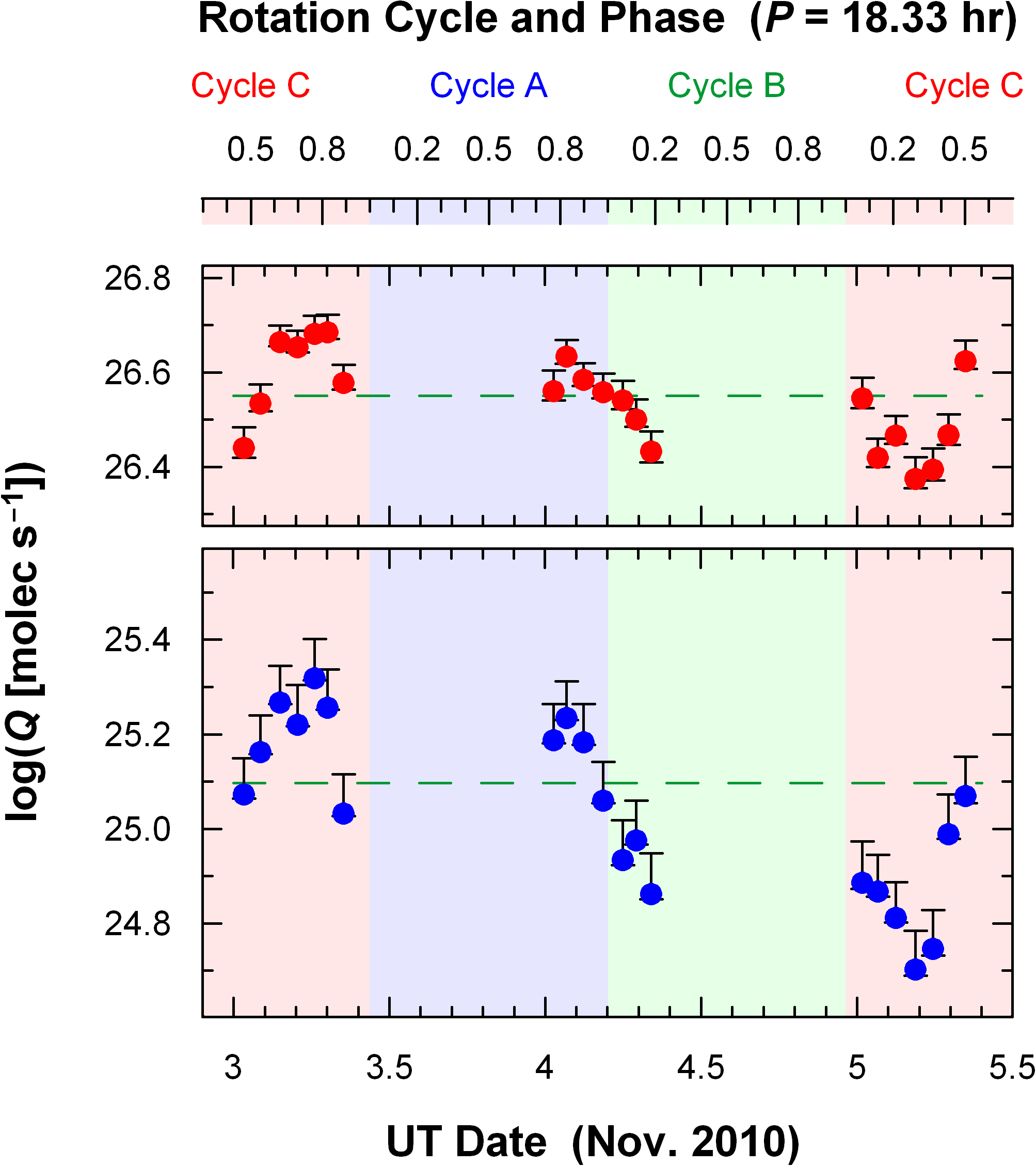}
\caption{Production-rate variability in CH$_3$OH (\emph{top panel})
and HCN (\emph{bottom panel}) presented in logarithmic scale. The
production rates $Q$ vary about the average values:
$3.5\times10^{26}$ molec~s$^{-1}$ for CH$_3$OH and
$1.25\times10^{25}$ molec~s$^{-1}$ for HCN (dashed lines). The
asymmetry of the error bars results from the uncertainty of the
telescope pointing (cf.~Paper~I), which surpasses the errors from
noise. For this reason, the uncertainties on the simultaneous
measurements of CH$_3$OH and HCN are partly correlated, unlike the
errors on subsequent data points. Systematic errors affecting only
the absolute levels were neglected (cf.~Section~\ref{Sec-Obs}) as
they do not change the shapes of the variability profiles. The
additional top axis shows the nucleus rotation phase and
\emph{three-cycle} component (cf.~Paper~I), and the latter is also
coded by the background color. The \emph{EPOXI} flyby occurred on
UT~2010 Nov.~4.5832, corresponding to phase 0.5 of
\emph{Cycle~B}.\label{Fig-EvolProdRates}}
\end{figure}

We observe well-defined variation in the instantaneous production
rates of both molecules (Fig.~\ref{Fig-EvolProdRates}), which we
previously noticed in HCN and connected with the rotation of the
nucleus (Paper~I). The two time series correlate remarkably well
although the amplitudes are different, reaching a factor of~4 for
HCN but only a factor of~2 for CH$_3$OH. However, the measured
amplitudes can differ from the real ones if some of our model
assumptions (Appendix~\ref{Ap-Model}) are strongly violated, in
particular (\emph{i}) the negligible optical depth, (\emph{ii}) the
constant rotational temperature, or (\emph{iii}) the outgassing
properties.

\placefigure{Fig-EvolProdRates}

A non-negligible optical depth would lead to an underestimation of
the production rates, most significant for the brightest lines.
Consequently, the amplitude of the brighter HCN would be reduced
compared to the fainter CH$_3$OH, and since we observe otherwise,
the real amplitude difference would have to be higher than measured.
Nevertheless, the optical depth is completely negligible in our
data, which is best evidenced by the hyperfine splitting of the
\mbox{HCN \emph{J}(3--2)} line (Fig.~\ref{Fig-AvgHCN}). We make use
of the fact that the relative intensities of the hyperfine line
components in a single molecule are established by fundamental
physics and do not depend on the excitation conditions or mechanisms
\citep[e.g.][]{Boc84}. Consequently, if a component ratio measured
in a molecular environment is equal to the theoretical value for a
single molecule and different from unity, it implies that these
components are optically thin. Taking advantage of the exceptionally
high signal-to-noise ratio in the average spectrum of HCN, we can
easily distinguish the $F=2-2$ hyperfine component and measure its
area (from $-3.45$ to $-1.5$~km~s$^{-1}$) separately from the blend
of the remaining components (from $-1.5$ to $+2.7$~km~s$^{-1}$). We
find the component ratio to be equal to $24.1^{+1.5}_{-1.3}$ which
is consistent with the theoretical ratio of 26.0 \citep[cf. Cologne
Database for Molecular
Spectroscopy\footnote{\url{http://www.astro.uni-koeln.de/cdms}};][]{Mul05}.
This means that even the brightest region of the line, which is
always first to saturate, is practically free of self-absorption,
and that the fainter lines of CH$_3$OH must be optically thin as
well.

Since the production rates were derived using the constant
rotational temperature, they can be naturally affected by the
measured temperature variations (Section~\ref{Sec-Trot}). From the
theoretical relation between these two quantities in
Fig.~\ref{Fig-ModelQVsTrot} we see that the derived production rate
depends on the temperature more strongly for CH$_3$OH than for HCN.
This implies that the variations of the line area could not have
been produced primarily by the varying temperature instead of the
production rate because the amplitude would be higher in CH$_3$OH
than in HCN. Consequently, it must be a real variability of the
production rate generating physical changes in the rotational
temperature rather than the temperature variations modulating our
derived production rates. However, the latter effect must also be
present to some extent. Bearing in mind the suspected positive
correlation of these two quantities in our data
(Fig.~\ref{Fig-TrotVsQ}), we expect that the derived production-rate
maxima are somewhat underestimated and the minima overestimated,
although this reasoning is limited to $T_\mathrm{rot} > 26$~K for
which both functions in Fig.~\ref{Fig-ModelQVsTrot} monotonically
increase (a lower $T_\mathrm{rot}$ was measured in only one data
point, spectrum~\#14, and is equal to 19~K). Consequently, the
observed amplitudes can be lower compared to the real ones, and
since the effect is stronger for CH$_3$OH than for HCN, it can
explain, at least to some degree, the different amplitudes measured
in our data.

In the above discussion we assumed that the temperature variation is
the same in HCN and CH$_3$OH. If, instead, the (unmeasured)
characteristic temperature applicable to our HCN data varied more
strongly, e.g. due to the smaller beam size (Table~\ref{Tab1}), the
range of the production-rate offsets from
Fig.~\ref{Fig-ModelQVsTrot} could even exceed the range for
CH$_3$OH. This would imply that the real production-rate amplitudes
differed more than inferred from our data because HCN would be
attenuated more strongly than CH$_3$OH. On the other hand, this
scenario still cannot explain the derived production-rate variations
as artifacts caused by the varying temperature (even though
correctly implying the measured amplitude relation) because in such
a case the two quantities would be inversely correlated while the
determined correlation is straight. Whereas the situation can be, in
principle, even more complex if e.g. the two variability profiles of
$T_\mathrm{rot}$ were totally different, we believe, rather, that
the beam-size effect is of secondary importance in this respect --
and hence the temperature in the data of both molecules is
comparable, as we cannot identify other effects that could
differentiate it significantly.

However, the different beam sizes can also affect the derived
production rates in other ways, especially when the outgassing
properties assumed in the model (Appendix~\ref{Ap-Model}) are
strongly violated. In fact, the outgassing of 103P is by no means in
steady state and isotropic, but rather dominated by rotating jets
and icy grains periodically injected into the coma (see the next
sections and references therein). In such a case, the observed
amplitude can be smaller than the real one, by a factor which
depends on the sublimation-time dispersion of the molecules
contributing to a single spectrum \citep[see e.g.][]{Biv07,Dra10}.
In this way, the lower amplitude of CH$_3$OH seems to be naturally
explained, given the larger beam size that effectively ``sees'' the
molecules from a broader range of nucleus rotation phases compared
to the smaller beam of HCN (Table~\ref{Tab1}). But this simple
reasoning fails to explain the remarkable temporal correlation of
both variability profiles, and also the fact that the maxima and
minima look relatively flat. Consequently, the timescale of
variations appears sufficiently long compared to the integration
time and escape time from the beam (both $\sim1$~hr) to ensure that
the outgassing properties are rather ``frozen'' on the spatial and
temporal scales characteristic of the individual CH$_3$OH and HCN
data points (Table~\ref{Tab1}). This may indicate that the two
diurnal amplitudes were indeed different. However, even if they were
the same, the different beam sizes would still differentiate them in
the same sense as observed if, for example, the minimum level was
produced by a constant background with uniform brightness
distribution, i.e. if the model assumption of a central source of
outgassing is not well satisfied. (Note that the production-rate
profiles of CH$_3$OH and HCN are still deformed by the outgassing
anisotropy, but both in a similar way.)

Last but not least, we note that the errors of telescope pointing
can also affect the derived production-rate profiles. Since the
effect is inversely correlated with the beam size, it affects HCN
more strongly than CH$_3$OH, but in the same sense, and therefore it
is consistent with their seeming temporal correlation and the
relation of their amplitudes. However, given that the pointing
corrections were always determined between the consecutive points in
the time series, this effect is practically incapable of generating
entire trends, introducing only small random deviations, and also
its expected magnitude is rather small which is reflected by the
derived error bars.

\begin{figure}
\epsscale{0.5} \plotone{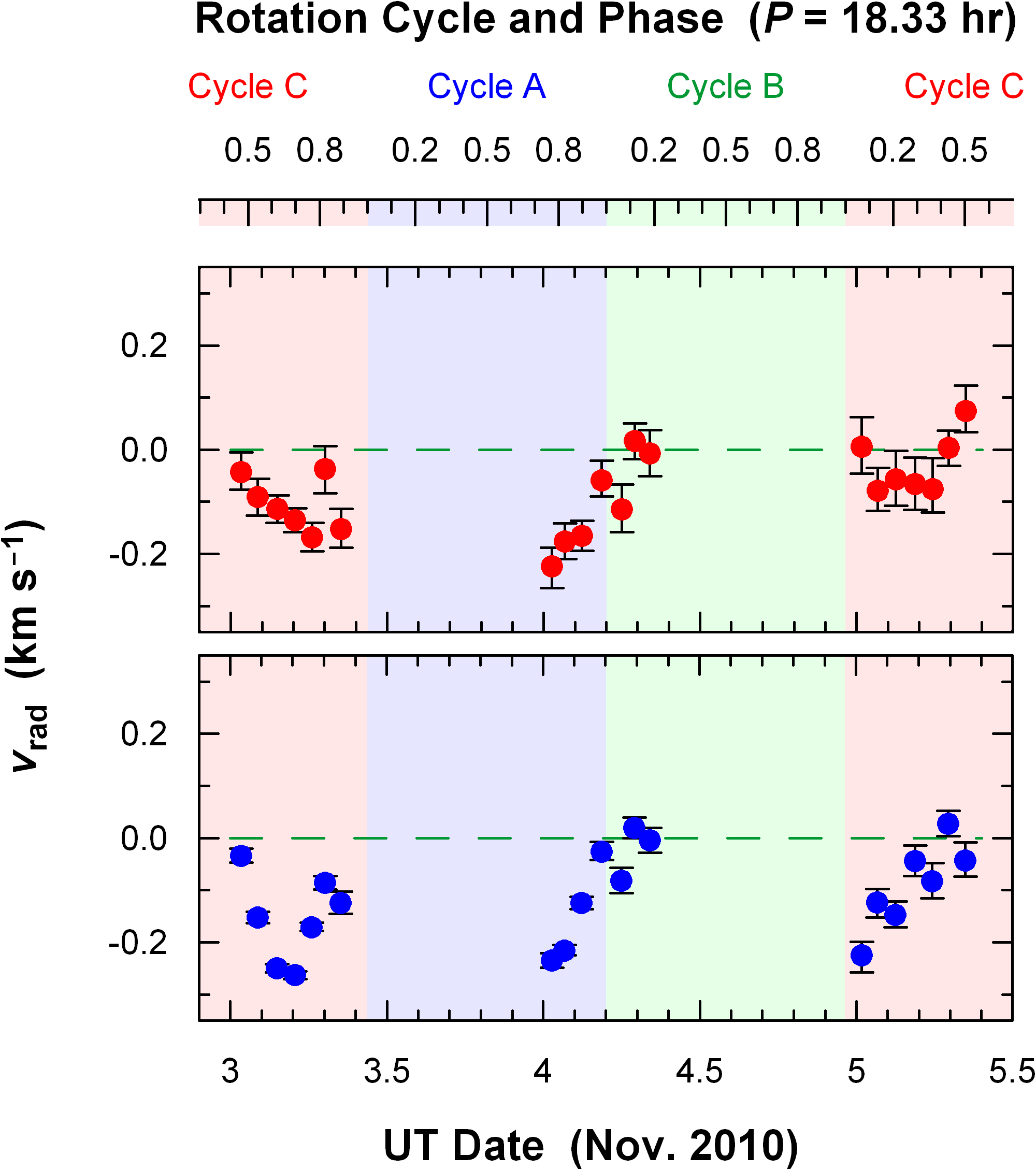}
\caption{Variability of the median radial velocity $v_\mathrm{rad}$
in CH$_3$OH (\emph{top panel}) and HCN (\emph{bottom panel}) about
the rest velocity (dashed lines). The uncertainty of the telescope
pointing is assumed to have a negligible influence on this parameter
and hence the errors result entirely from noise (Paper~I). The
additional top axis shows the nucleus rotation phase and
\emph{three-cycle} component (cf.~Paper~I), and the latter is also
coded by the background color. The \emph{EPOXI} flyby occurred on
UT~2010 Nov.~4.5832, corresponding to phase 0.5 of
\emph{Cycle~B}.\label{Fig-EvolVrad}}
\end{figure}

\begin{figure}
\epsscale{0.5} \plotone{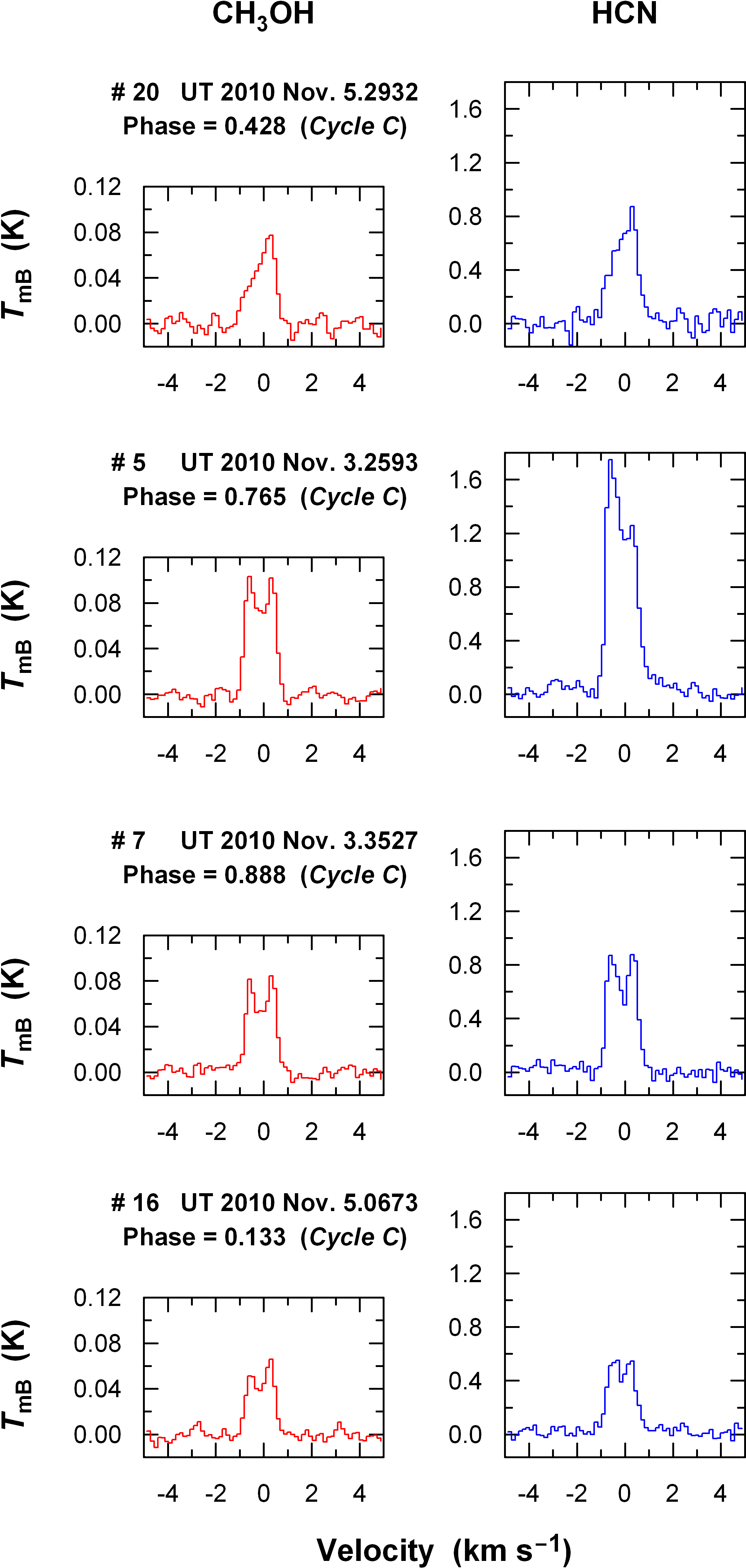}
\caption{Examples of the line profiles of CH$_3$OH (\emph{left
panels}) and HCN (\emph{right panels}) illustrating the temporal
variations as the nucleus rotates (from top to bottom). The former
are the mean profiles from the observed five lines whereas the
latter are the profiles of the \mbox{\textit{J}(3--2)} transition
(Section~\ref{Sec-Obs} and Table~\ref{Tab1}). We selected four
representative rotation phases from two consecutive \emph{Cycles~C}
(cf.~Paper~I) observed on UT~2010 Nov.~3 (\emph{middle panels}) and
Nov.~5 (\emph{top} and \emph{bottom panels}). The complete dataset
is presented in Fig.~\ref{Fig-AllLineShapes}
(Appendix~\ref{Ap-Data}).\label{Fig-ExLineShapes}}
\end{figure}

In order to fully understand the characteristics of the two
production-rate profiles, we need to take into account the velocity
information naturally contained in the lineshapes.
Figure~\ref{Fig-EvolVrad} shows the variability of the median radial
gas-flow velocity $v_\mathrm{rad}$, which has been assumed equal to
the median line velocity $v_0$ (Appendix~\ref{Ap-Model}). The
behavior is also well defined and consistent with the rotational
periodicity of the nucleus. We see that the line-of-sight kinematics
were noticeably different for the two molecules, in particular, HCN
drifts towards deeply blueshifted velocities around the phases of
maximum activity, while the velocities of CH$_3$OH change less. In
Fig.~\ref{Fig-ExLineShapes} we show examples of the individual
lineshapes. A close-up look reveals that HCN has a strong
blueshifted component at the active phases, while CH$_3$OH looks
symmetric with two distinct peaks. The difference vanishes at the
interim and quiescent phases, at which the lineshapes look
remarkably similar to each other -- featuring two symmetric peaks
which resemble the lineshape of CH$_3$OH at the active phases
(although the redshifted peak occasionally dominates the spectra of
both molecules). The implication of these differences is that the
sources of the two molecules must have differed in some way. We
further explore this inference in the next section.

\placefigure{Fig-EvolVrad}

\placefigure{Fig-ExLineShapes}

\subsection{Interpretation of the Lineshapes}\label{Sec-Sources-LineShapes}

\subsubsection{Basics of Lineshape Interpretation}\label{Sec-Sources-LineShapes-Basics}

Since the characteristic gas-flow velocity in comets 1~AU from the
Sun is $\sim1$~km~s$^{-1}$ \citep[e.g.][]{Com04}, our spectra are
fully ``velocity resolved'' given that the spectral resolution is
0.15~km~s$^{-1}$ (Section~\ref{Sec-Obs}). The lineshapes in such
spectra are primarily controlled by the Doppler effect. Other
effects, including optical depth and hyperfine structure, can also
theoretically influence the lineshapes but are negligible in our
data. The velocity-calibrated spectra can be hence identified as
histograms, showing how the molecules observed within the beam are
distributed in radial velocity, weighted by the \emph{emission
coefficient} (which is a function of the excitation conditions in
different regions of the coma).

\begin{figure}
\epsscale{0.5} \plotone{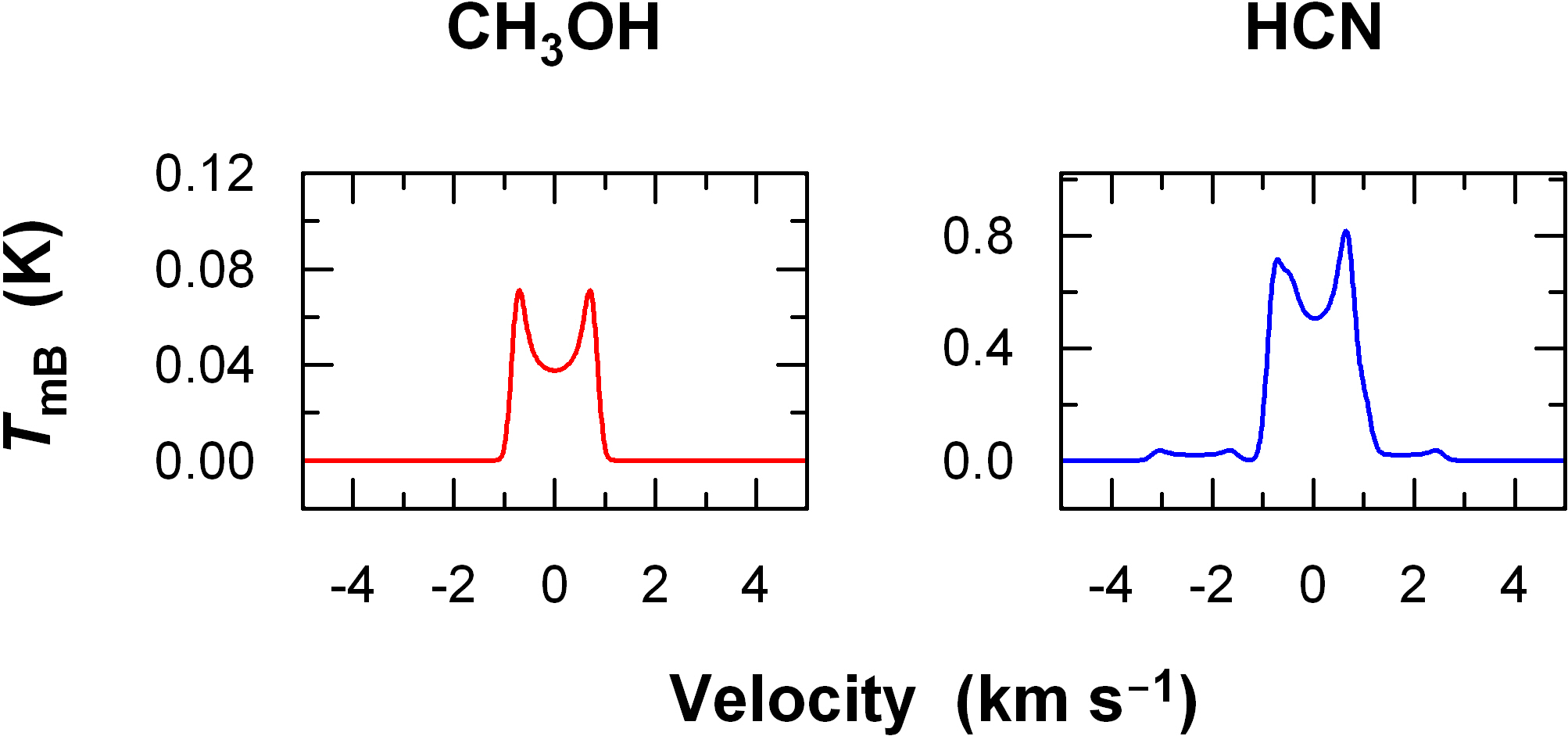}
\caption{Simulated spectra of CH$_3$OH (\emph{left panel}) and HCN
(\emph{right panel}) for steady-state isotropic outgassing from a
central source with the physical and geometric conditions compatible
with our data. We used the production rates of $3.5\times10^{26}$
and $1.25\times10^{25}$ molec~s$^{-1}$, respectively
(Section~\ref{Sec-Sources-ProdRates}), a constant gas-flow velocity
of 0.8~km~s$^{-1}$ (Appendix~\ref{Ap-Model}), and LTE at 47~K
(Section~\ref{Sec-Trot}). The beam sizes and the molecular constants
were used the same as given in Table~\ref{Tab1}
(Section~\ref{Sec-Obs}). The HCN profile additionally accounts for
the hyperfine structure taken from the Cologne Database for
Molecular Spectroscopy \citep[][]{Mul05}, which introduces the
asymmetry. This information is currently unavailable for the
observed lines of CH$_3$OH, and hence the mean model profile is
symmetric.\label{Fig-ModelSpectra}}
\end{figure}

This distribution depends on the directions in which the molecules
travel. In particular, molecules traveling along the line of sight
are either maximally redshifted or blueshifted, while molecules
traveling normal to the line of sight are observed at zero velocity
(corresponding to their rest frequencies). If the molecules are
ejected isotropically at a constant speed, then the emission line
should feature two peaks symmetrically located about 0~km~s$^{-1}$,
with radial velocities equal to the positive and negative values of
the gas-flow velocity (see model examples in
Fig.~\ref{Fig-ModelSpectra}). This is because most molecules in the
beam travel either towards or away from the observer along the line
of sight. Molecules traveling perpendicular to the line of sight
leave the beam fastest producing a central minimum in the line
profile. Moreover, real gas always has a ``thermal'' velocity
component that blurs the lineshape. This effect is of secondary
importance, however. For example, the thermal speed of CH$_3$OH at
47~K (Section~\ref{Sec-Trot}) is only 0.2~km~s$^{-1}$, small
compared to the bulk gas-flow velocity of $\sim1$~km~s$^{-1}$.

\placefigure{Fig-ModelSpectra}

The travel directions of the molecules, at least in the inner coma,
can be identified with the directions in which they were ejected,
and therefore, in the discussion below, we directly link the
observed spectral features with the characteristics of 103P's
outgassing. At this stage we only aim at interpreting the most
obvious spectral features and only in a qualitative manner. A more
in-depth analysis requires a detailed lineshape modeling
\citep[cf.][]{Dra09}, which is presently in preparation.

\subsubsection{Evolution of the Lineshapes}\label{Sec-Sources-LineShapes-Evol}

The evolution of the lineshapes (see Fig.~\ref{Fig-ExLineShapes})
can be grouped in three distinct ranges of the nucleus rotation
phase:

\begin{list}{\labelitemi}{\leftmargin=1em}

\item Phases 0.4--0.5 (spectra \#1 and \#20--21): the \emph{redshifted jet}.

The line profiles of CH$_3$OH and HCN look very similar. They
feature a well-defined redshifted peak at about $+0.3$~km~s$^{-1}$,
which we identify with a jet directed away from the Earth (as
projected onto the line of sight). The gas in the jet was abundant
in both CH$_3$OH and HCN, and must originate from a vent (or a group
of vents) on the nucleus, which was active only at these particular
rotation phases. We refer to this feature as to the \emph{redshifted
jet}.

\item Phases 0.5--1.0 (spectra \#2--7, \#8--11, and \#21): the \emph{blueshifted jet}.

Essentially half of the rotation cycle, between phases 0.5 and 1.0,
is dominated by the appearance of a second jet (this time
blueshifted) and associated phenomena:

\begin{itemize}

\item[--] The first signatures are visible in HCN around phase 0.5.
The HCN lineshape starts showing a blueshifted peak, which is
visible simultaneously with the redshifted peak from the earlier jet
(spectrum \#21, and between \#1 and \#2). At the same time, CH$_3$OH
still shows only the redshifted peak.

\item[--] Around phase 0.55 (spectrum \#2) the blueshifted peak
starts dominating the HCN spectrum, while CH$_3$OH becomes symmetric
with two distinct peaks.

\item[--] Between phases 0.6 and 0.8 (spectra \#3--5 and \#8--9) both
molecules rapidly brighten while their lineshapes continue to
starkly differ from each other: HCN is totally dominated by a
blueshifted peak near $-0.6$~km~s$^{-1}$ and CH$_3$OH shows two
rather equal peaks located symmetrically about $-0.15$~km~s$^{-1}$
(one close to $-0.6$~km~s$^{-1}$ and the other one near
$+0.3$~km~s$^{-1}$).

\item[--] After reaching the maximum brightness around phase 0.8,
both molecules start fading until about phase 1.0 (spectra \#6--7
and \#10--11). This changeover is associated with a dramatic
transformation of the HCN lineshape. It starts showing two symmetric
peaks, just like CH$_3$OH that remains unchanged, and so the
lineshapes of the two molecules look very similar while the
brightness decreases.

\end{itemize}

The lineshapes dominated by the blueshifted peak at about
$-0.6$~km~s$^{-1}$ can be identified with a jet originating from a
vent (or a group of vents) on the nucleus, which was active only at
these particular rotation phases. The gas was produced Earthward as
projected onto the line of sight, and hence we refer to this feature
as to the \emph{blueshifted jet}. On the other hand, the symmetric
double-peak lineshapes, blueshifted by $\sim0.15$~km~s$^{-1}$, can
be most easily interpreted as created by an isotropic source, also
moving Earthward (in the same sense as above) but about four times
slower than the gas in the jet.

We associate the isotropic source with ice particles in the coma of
103P. The \emph{EPOXI} flyby revealed large icy grains close to the
nucleus, up to 10--20~cm in radius \citep{AHe11a}, and also the
Arecibo radar detected centimeter to decimeter grains \citep{Har11}.
However, the velocities of these chunks, mostly below
$0.5$~m~s$^{-1}$ in the \emph{EPOXI} data and of the order of few to
tens of meters per second as measured at Arecibo, are too low to
cause a measurable line shift in our spectra. Instead, we believe
that the total sublimation area of the grains was controlled by the
finest particles \citep[cf.][]{AHe11a}, which were ejected at higher
speeds and further accelerated by the gas in the coma. For example,
taking the grain velocity as a function of size as derived for this
comet from the radar data \citep{Har11}, we obtain $0.2$~km~s$^{-1}$
for particles $\sim0.5$~mm in diameter. Such grains would be
isothermal and could survive for perhaps an hour at this
heliocentric distance, which altogether makes them excellent
candidates to explain the symmetric double-peak line profiles and
their rapid temporal evolution.

The grains appear to be injected into the coma and further
accelerated by the gas in the \emph{blueshifted jet}. Consequently,
the velocity projection onto the line of sight was basically the
same for the gas and ice in the jet, and hence we identify their
Doppler-shift ratio of $\sim4$ with the actual gas-to-grains speed
ratio. We also suppose that the projection effect was not very
strong for this jet (see further in Section~\ref{Sec-Discussion}),
implying that the submm grains traveling at $\sim0.2$~km~s$^{-1}$
could also plausibly explain the amount of the systematic blueshift
in the double-peak line profiles.

The spectra show that the gas originating directly at the vent was
rich in HCN but not in CH$_3$OH. In contrast, the excavated ice
particles sublimated isotropically and carried abundant HCN and
CH$_3$OH in proportions comparable to the gas in the
\emph{redshifted jet} observed earlier. Consequently, at these
rotation phases, the observed HCN sublimated both from the icy
grains and from the active vent, while the observed CH$_3$OH
sublimated only from the ice particles. The composition of the
grains appears then dramatically different from the composition of
the gas in the \emph{blueshifted jet} which carried them away from
the nucleus.

The above conclusions are consistent with the temporal evolution of
both line profiles. At the onset of the \emph{blueshifted jet}, the
observed HCN coma was dominated by the gas in this jet, and
therefore the HCN line displays the strongly blueshifted peak.
However, at the same time, the observed coma of CH$_3$OH was
dominated by the gas sublimating from the first grains in the jet,
and therefore CH$_3$OH started brightening with some delay,
presenting the subtly blueshifted symmetric double-peak lineshape.
With time, the grains became more abundant in the coma, and also the
outgassing rate from the vent increased, injecting more gas and ice.
This corresponds to the rapid brightening visible for both molecules
and explains why their profiles continued to differ so much:
CH$_3$OH, sublimating only from the grains (isotropically),
preserved the two rather equal peaks, while HCN, produced
additionally in the active vent (anisotropically), continued to be
dominated by the single peak. At some point, the sublimation from
the vent declined, but the ice particles emitted earlier persisted,
and that is when the lineshapes of CH$_3$OH and HCN became
strikingly similar to each other, showing two symmetric peaks. While
the grains continued to lose their total cross-section and move away
from the beam center, the lines kept fading although their shapes
remained unchanged.

\item Phases 0.0--0.4 (spectra \#12--19): the quiescent rotation phases.

The CH$_3$OH and HCN lineshapes look rather similar. In most of the
spectra, they show the subtly blueshifted symmetric profiles, which
we have earlier identified with the isothermal icy grains. The
brightness of the lines stabilized at some minimum level, i.e.
neither strongly declined with the increasing rotation phase nor
rose up with the increasing count of the rotation cycles. This leads
us to the conclusion, that the total sublimation area of these
particles was fairly constant on a rotation-cycle timescale. Some of
the grains perhaps originated from the earlier rotation phases, but
this source must have been quickly decaying after deactivation of
the \emph{blueshifted jet}, because of the mass loss from
sublimation and decreasing telescope sensitivity as the particles
moved away from the beam center. The grains must have been
replenished by other sources, perhaps including weaker vents
(apparently incapable of generating strong jet signatures in our
spectra) or fragmentation of larger chunks with subsequent
acceleration of the created particles by the gas in the inner coma.

\end{list}

\subsubsection{Additional Remarks}\label{Sec-Sources-LineShapes-Remarks}

The proposed scenario naturally explains the observed
characteristics of the HCN and CH$_3$OH production-rate profiles
(Section~\ref{Sec-Sources-ProdRates}). Bearing in mind that the
observed lines reacted almost instantly to the changes in comet's
activity (cf. Section~\ref{Sec-Sources-ProdRates}), the remarkable
correlation between the two molecules is understandable given that
their maxima are controlled by the same source of activity (the
\emph{blueshifted jet} containing gas and ice) and likewise for the
minima (controlled by the background of icy grains). Simultaneously,
the difference between the production-rate amplitudes is explained
by the HCN excess at the phases of maximum activity, causing a
larger range of variation compared to CH$_3$OH. Other scenarios,
which are discussed in Section~\ref{Sec-Sources-ProdRates}, cannot
be accepted because they fail to explain the observed differences in
the lineshapes. Specifically:
\begin{list}{\labelitemi}{\leftmargin=1em}

\item The varying optical depth can, in principle, produce temporal changes in
the line profiles, but we concluded that both molecules were
optically thin in our data.

\item The variations in the rotational temperature are rather difficult to
connect with the lineshapes.

\item Potential problems with pointing are unlikely to generate
systematic trends or, for the still stronger reason, periodic
trends.

\item In the suspected situation in which the production-rate
amplitudes are differentiated by the different beam sizes (due to
background) but the sources of CH$_3$OH and HCN in the comet are the
same, CH$_3$OH would also display a blueshifted component at the
phases dominated by the \emph{blueshifted jet} (albeit somewhat
fainter than in HCN) but we do not see it in our data.

\item We also note that the line profiles of CH$_3$OH and HCN
naturally differ due to the presence of hyperfine splitting, but
this difference is ``static'' for optically-thin lines and rather
small (Fig.~\ref{Fig-ModelSpectra}).

\end{list}

While none of the above mechanisms can be a plausible alternative to
the postulated difference in the sources of CH$_3$OH and HCN, some
of them could possibly be held responsible for small deviations from
our preferred scenario, which naturally exist in our data. Small
inconsistencies could also be attributed to the weaker outgassing
sources observed by \emph{EPOXI} \citep{AHe11a}, to the non-trivial
dynamics of the comet's coma, and also to the reported excitation of
the nucleus rotation state \citep[cf.~Paper~I;
also][]{AHe11a,Kni11,Sam11,Wan12} which we explore in the next
section. When attempting to identify the smallest inconsistencies,
one should also keep in mind some unavoidable limitations of our
dataset, such as the finite signal-to-noise ratio and baseline
reliability, and the limited pointing accuracy of the telescope, as
well as the rotation-phase errors possibly caused by the limited
knowledge of the rotational dynamics of the nucleus.

\subsubsection{Differences between Rotation Cycles}\label{Sec-Sources-LineShapes-Diffs}

It is interesting to note that some well-defined deviations from the
picture presented in Section~\ref{Sec-Sources-LineShapes-Evol} can
be plausibly associated with the excitation of 103P's rotation
state. They also agree remarkably well with the \emph{three-cycle}
scenario that we introduced in Paper~I to approximate repeatability
in rotation-modulated data of this comet. Specifically:

\begin{list}{\labelitemi}{\leftmargin=1em}

\item Phases 0.1--0.2 (spectra \#13--14) on UT~Nov.~4
(\emph{Cycle~B}) show that the lines of CH$_3$OH and HCN are
consistently dominated by the redshifted peak, resembling the
lineshapes previously identified with the \emph{redshifted jet} at
phases 0.4--0.5 (Section~\ref{Sec-Sources-LineShapes-Evol}).
However, on UT~Nov.~5 (\emph{Cycle~C}) the lines at phases 0.1--0.2
(spectra \#16--17) are symmetric and fainter than on UT~Nov.~4
(\emph{Cycle~B}).

We suppose that the redshifted lineshapes, appearing at the two
separate phase ranges, might have been produced by the same
\emph{redshifted jet} identified earlier. Because of the excitation
of 103P's rotation and the likely circumpolar origin of the
\emph{redshifted jet} (see further in Section~\ref{Sec-Discussion}),
the parent vent might have been activated by sunlight at phases
0.4--0.5 during \emph{Cycle~C}, but earlier -- at phases 0.1--0.2 --
during \emph{Cycle~B}. This interpretation is consistent with the
fact that in both cases the lineshapes of CH$_3$OH and HCN look very
similar (unlike at the phases dominated by the \emph{blueshifted
jet}) and is additionally confirmed by the HCN data from UT~Nov.~2
(see further in this paragraph).

\item The excitation of the rotation state is also likely
responsible for the fact that at the rotation phase 0.82 of
\emph{Cycle~C} (spectrum \#6 obtained on UT~Nov.~3) the lineshapes
are evidently more evolved than at the same phase of \emph{Cycle~A}
(spectrum \#9 obtained on UT~Nov.~4). Given the remarkable
coincidence in phase of the \emph{redshifted jet} observed during
two consecutive \emph{Cycles~C}
(Section~\ref{Sec-Sources-LineShapes-Evol}), the above inconsistency
cannot be removed by simply adjusting the \mbox{18.33-hr} rotation
period.

\item The spectra of HCN from UT~Nov.~2 (formally excluded from
the analysis because no counterpart spectra of CH$_3$OH were taken),
covering phases 0.2--0.6 of \emph{Cycle~B}, show that the
\emph{redshifted} and \emph{blueshifted jets} appeared at relatively
earlier rotation phases than in \emph{Cycle~C} observed on UT~Nov.~3
and~5. This behavior is consistent with the rotational
production-rate profiles of HCN from a much bigger dataset presented
in Paper~I.

On UT~Nov.~2 the \emph{redshifted jet} is visible at phases
0.2--0.25, which agrees with the short \emph{Cycle~B} data from
UT~Nov.~4 (covering phases 0.05--0.2) in which we have identified
this jet at phases 0.1--0.2 (see earlier in this paragraph).
Combining these two pieces of \emph{Cycle~B} data we conclude that
the \emph{redshifted jet} was present at phases 0.1--0.25, with a
subtle local minimum at phase 0.2.

Moreover, in \emph{Cycle~B} the blueshifted peak looks fainter than
in \emph{Cycles~C} and~\emph{A}, which may indicate that the
\emph{blueshifted jet} was physically weaker (e.g. due to different
insolation), although other explanations may be possible as well
(e.g. a different projection effect). Whatever the cause, in
\emph{Cycle~B} the \emph{blueshifted jet} triggers a comparable
quantity of icy grains as in \emph{Cycles~C} and~\emph{A} and,
consequently, at the phases of maximum brightness the HCN line
features a symmetric double-peak profile.

\end{list}

\section{Discussion}\label{Sec-Discussion}

The velocity-resolved spectral time series of CH$_3$OH and HCN
reveal a complex outgassing portrait of 103P. We identify outgassing
in the form of at least two jets -- the \emph{redshifted jet} and
the \emph{blueshifted jet} -- expanding in the opposite directions
as projected onto the line of sight and having seemingly different
compositions. The \emph{blueshifted jet} appears as a strong source
of HCN gas and of icy grains, but not of CH$_3$OH gas. The
composition of the grains is, however, dramatically different from
the composition of the gas in this jet because, in addition to HCN,
they also contain large amounts of CH$_3$OH. Surprisingly, in terms
of composition the grains are very similar to the gas in the
\emph{redshifted jet}, which is also rich in HCN and CH$_3$OH
although it does not carry much ice. Consequently, the origin of the
observed CH$_3$OH and HCN can be summarized as follows:
\begin{list}{\labelitemi}{\leftmargin=1em}
\item CH$_3$OH:
  \begin{itemize}
  \item[--] anisotropic sublimation from the active vent that produced the \emph{redshifted jet},
  \item[--] isotropic sublimation from the icy grains carried away primarily in the \emph{blueshifted
  jet}, but not from the vent itself (at least not at any measurable level).
  \end{itemize}
\item HCN:
  \begin{itemize}
  \item[--] anisotropic sublimation from the active vent that produced the \emph{redshifted jet},
  \item[--] anisotropic sublimation from the active vent that produced the \emph{blueshifted jet},
  \item[--] isotropic sublimation from the icy grains carried away primarily in the \emph{blueshifted jet}.
  \end{itemize}
\end{list}

It is interesting to note that, at the comet's phase angle $\sim
60\degr$ (Section~\ref{Sec-Obs}), the Sunward direction projects as
blueshifted and most of the illuminated part of the nucleus would
emit material in the blueshifted directions. However, a smaller but
still substantial fraction of the illuminated part would emit
material in the redshifted directions. For this reason, it is
entirely possible that both the vent producing the \emph{blueshifted
jet} and the vent producing the \emph{redshifted jet} were
illuminated while being active. However, we suppose that the
\emph{redshifted jet} originate from a vent close to the region of
the polar night, which was illuminated only during a short fraction
of the rotation cycle. This would naturally explain its short
duration and opposite radial-velocity component compared to the
\emph{blueshifted jet}. Supposing that the gas-flow velocity was
comparable in the two jets, perhaps close to 0.8~km~s$^{-1}$
(Appendix~\ref{Ap-Model}), we take into account the difference in
velocity offsets of the line peaks ($+0.3$~km~s$^{-1}$ for the
\emph{redshifted jet} vs. $-0.6$~km~s$^{-1}$ for the
\emph{blueshifted jet}) and conclude different significance of the
projection effect -- implying that the \emph{blueshifted jet} was
emitted in a direction close to the line of sight, while the
\emph{redshifted jet} was relatively closer to the sky plane. We
note that the fact that the two seemingly different jets appeared at
consecutive rotation phases makes them hard to distinguish in the
time series of the total production rate
(Fig.~\ref{Fig-EvolProdRates}). However, expanding in the opposite
directions (as projected onto the line of sight), they clearly
manifest themselves as separate features in the velocity-resolved
line profiles.

The properties of these two jets resemble very much the CN features
observed around the same time in narrowband optical images
\citep{Kni11,Sam11,Wan12}. In particular, the images of
\citet{Wan12} -- obtained nearly simultaneously with our IRAM~30~m
data -- are dominated by structures with low projected sky-plane
velocities, hence expanding close to the line-of-sight direction
(the \emph{slow CN feature}). We tentatively associate these
features with our \emph{blueshifted jet}, given the dominating role
of the \emph{blueshifted jet} in our data and also its strong
Doppler shift. However, a small part of the dataset of \citet{Wan12}
shows a much faster feature, expanding in the \mbox{S--SW} direction
(the \emph{fast CN feature}) probably not far from the sky plane.
Interestingly, \citet{Kni11} and \citet{Sam11} report that the
\mbox{S--SW} feature became first visible only in October.
Therefore, we suppose that it is the same structure as our
\emph{redshifted jet}, having a relatively weak Doppler shift, and
probably just emerging from the polar night.

We have earlier established (Paper~I) that the variability of the
HCN production rate was in phase with the brightness variation in
CO$_2$ and H$_2$O observed by \emph{EPOXI} \citep{AHe11a}.
Therefore, CH$_3$OH also correlates with all these molecules. But
even though the variations of the four molecules were in phase, the
spatially resolved molecular observations from \emph{EPOXI} revealed
different reservoirs of CO$_2$ and H$_2$O, and our own observations
show differences between HCN and CH$_3$OH. Interestingly, however,
the variability amplitudes\footnote{Due to the excitation of the
rotation state, the pattern of variability repeats best every three
rotation cycles \citep[Paper~I; also][]{AHe11a}. We observed the
minimum and maximum levels during different rotation cycles but they
correspond to the same \emph{\mbox{three-cycle}} component
(\emph{Cycle~C}). Nevertheless, due to the residual differences
between the same \emph{three-cycle} component observed at different
times, the total amplitudes of HCN and CH$_3$OH should be
interpreted with some caution.} of HCN and CH$_3$OH, reaching a
factor of~4 and a factor of~2, respectively, are the same as
measured for CO$_2$ and H$_2$O. An intriguing hypothesis arises
then, that perhaps HCN is spatially correlated in the nucleus with
CO$_2$ and CH$_3$OH with H$_2$O. Consequently, we suppose that the
\emph{redshifted jet} (tentatively associated with the \emph{fast CN
feature}) was observed by \emph{EPOXI} as the \emph{water jet},
while the \emph{blueshifted jet} (tentatively associated with the
\emph{slow CN feature}) was observed by the spacecraft as the
\emph{carbon dioxide jet}. Moreover, the \emph{carbon dioxide jet}
is the main supplier of icy grains in the \emph{EPOXI} data
\citep{AHe11a} and the same is true for the \emph{blueshifted jet}
in our data.

At this stage it is difficult to accurately quantify how much gas
originated directly at the nucleus and how much was produced from
the grains in the coma. Nevertheless, taking into account that in
about half of the HCN spectra, and in most of the CH$_3$OH ones, the
lines have two symmetric peaks (which we associate with the grains),
we can very crudely estimate that probably about half of the HCN
molecules, and most of the CH$_3$OH ones, sublimated from the
grains. This conclusion is in agreement with the statement of
\cite{AHe11b} based on the \emph{EPOXI} data. However, we also note
that in \emph{Cycle~B} covered on UT~Nov.~2 (just one
\emph{three-cycle} before the \emph{EPOXI} flyby, which occurred at
the middle of the next \emph{Cycle~B}), the \emph{blueshifted jet}
appeared weaker than in \emph{Cycles~A} and~\emph{C}, but excavated
a comparable amount of ice. Consequently, at phase 0.5 of this
cycle, corresponding to the situation at the encounter, the HCN
spectra are totally dominated by the icy grains. This may indicate
that \emph{EPOXI} visited 103P when the abundance of ice in the coma
was exceptionally high compared to the gas directly sublimating from
the nucleus and hence the information inferred from the flyby may
not represent the typical behavior of this comet.

It is now our ongoing effort to model the entire set of line
profiles in a non-steady-state anisotropic fashion
\citep[cf.][]{Dra09}, simultaneously with the image time series of
CN \citep{Wan12}. We wish to take advantage of the fact that both
types of data contain the velocity information in orthogonal
dimensions and hence are fully complementary. The result will be
used to constrain a 3D outgassing portrait and retrieve the source
locations of CH$_3$OH and HCN, which will be readily comparable with
the spatially resolved observations of CO$_2$ and H$_2$O from
\emph{EPOXI}. This will enable us to verify the suspected identities
of the various features emerging from the different datasets, which
are only hypothetical at this stage. We will also better understand
the sublimation dependence on insolation, quantify the gas
contribution from the icy grains, and obtain realistic,
time-dependent absolute production rates and gas-flow velocities.

The compositional heterogeneity with respect to CO$_2$ and H$_2$O
can be possibly explained by the different characteristic
sublimation temperatures: 72~K and 152~K, respectively
\citep{Yam85}. This should compositionally decouple these two
molecules both in the protosolar environment (different locations of
the \emph{snowlines}) and also later, in the thermally processed
ices of Jupiter-family comets \citep{Gui11}. In contrast, HCN and
CH$_3$OH have practically the same sublimation temperatures, 95~K
and 99~K, respectively \citep{Yam85}, and therefore the observed
heterogeneity may suggest that they underwent different condensation
processes. Perhaps HCN can at least partly escape from water ice,
while CH$_3$OH is fully incorporated in it -- either in the form of
clathrate-hydrates or as trapped gas within the (amorphous) ice
matrix \citep[e.g.][]{Pri04}. If true, it seems entirely possible
that a part of 103P's nucleus has been more heated in the past and
therefore lost its volatiles except for water and the molecules
incorporated in water ice (CH$_3$OH, HCN, ...), and that is where
the \emph{redshifted jet} originates from. At the same time, another
part can be more primordial and dominated by free volatiles (CO$_2$,
HCN, ...) rather than H$_2$O and the water-bonded elements, and that
is where the \emph{blueshifted jet} is formed. However, water and
the molecules incorporated in water ice are emitted from this part
in the form of solid icy grains, and therefore we suppose that they
also directly sublimate from this area, but at rates so much lower
compared to the free volatiles that they remain undetected.

If this scenario is true, the nucleus of 103P might have been born
as a body in which the molecules were uniformly mixed in a bulk
sense. In that case we would expect that the \emph{redshifted jet}
was much stronger in the past, when powered by the free volatiles as
currently observed in the \emph{blueshifted jet}. However, in the
course of time the thermal evolution of the body reversed the jet
strengths. It is suggestive that we currently observe the
\emph{blueshifted jet} on its way to completely lose the free
volatiles and become compositionally similar to the evolved
\emph{redshifted jet} but much weaker. This process, if sufficiently
fast, may be responsible for the significant decline in activity of
103P compared to the previous return \citep[cf.][albeit measured
through a proxy for water]{Com11}. Last but not least, given the
above evolutionary constraints, it is more appropriate to
characterize the \emph{redshifted jet} as ``HCN-depleted'' rather
than the \emph{blueshifted jet} as ``CH$_3$OH-depleted'', although
from the observational point of view it seems counterintuitive at
first glance.

Our results demonstrate the scientific potential of dense time
series of velocity-resolved spectra taken at millimeter wavelengths.
We found large-scale heterogeneity of 103P's nucleus with respect to
HCN and CH$_3$OH, which was not reported by two groups observing
independently in the infrared. We suspect that the dense time series
of \citet{Del11} is too incomplete to show the compositional
differences between different parts of the nucleus, as it covers
only 17\% of the rotation cycle in HCN (two exposures) overlapping
with 28\% coverage in CH$_3$OH (five exposures). The sparsely
sampled data of \citet{Mum11} may suffer likewise. Moreover, insight
into the line-of-sight kinematics, naturally available for comets
via millimeter spectroscopy, is still unreachable in IR. On the
other hand, the IR observations are complementary to those in the
millimeter wavelength regime by providing spatial profiles along the
slit, which are more difficult to obtain using single-dish
millimeter spectroscopy. Both groups observing 103P in the infrared
report similar distributions of H$_2$O and CH$_3$OH, while HCN
differed to some extent -- the characteristics that our work fully
supports.

\section{Summary}\label{Sec-Summary}

We observed CH$_3$OH and HCN in comet \mbox{103P/Hartley 2} using
the IRAM~30~m telescope. Velocity-resolved spectra taken between
UT~2010 Nov.~3.0 and~5.4 at a spectral resolution $2\times10^6$ show
strong variability of the production rate, median radial velocity,
detailed lineshape, and -- in the case of CH$_3$OH -- of the
rotational temperature. We associate the observed variations with
the properties of different regions of the nucleus successively
exposed to sunlight over the course of its rotation. Our results can
be summarized as follows:
\begin{enumerate}

\item We identify three distinct outgassing components in
velocity-resolved spectral line data. There are two jets with
opposite radial velocities (the \emph{redshifted jet} and the
\emph{blueshifted jet}) and an isotropic component evidently
produced by sublimation from submillimeter icy grains. The latter
are injected into the coma primarily through the \emph{blueshifted
jet}.

\item The nucleus of 103P is globally heterogeneous with respect to
CH$_3$OH and HCN. Collimated flows of these two molecules are
present in the \emph{redshifted jet}, but only HCN flow is detected
in the \emph{blueshifted jet}. Both molecules are also detected in
the icy grains in proportions comparable to those in the
\emph{redshifted jet}.

\item HCN is probably partly incorporated in water ice (either in the
form of clathrate-hydrates or as trapped gas within the amorphous
ice matrix) but also exists unbonded to water, while CH$_3$OH is
mostly or fully trapped.

\item The vent producing the \emph{redshifted jet} appears thermally
evolved (depleted in free volatiles). We suppose that it was located
close to the region of the polar night, and tentatively link it with
the \emph{fast CN feature} visible from the ground and the
\emph{water jet} observed by \emph{EPOXI}. The vent producing the
\emph{blueshifted jet} appears more primordial (rich in free
volatiles), and we tentatively link it with the \emph{slow CN
feature} visible from the ground and the \emph{carbon dioxide jet}
observed by the spacecraft.

\item The variations in both molecules show small but obvious
deviations from strict periodicity which are consistent with the
\emph{three-cycle} repeatability pattern. We interpret this as
another indication of the excitation of the nucleus rotation state.

\item The rotational temperature of CH$_3$OH varies strongly and is
loosely correlated with the varying production rate.

\item The average rotational temperature is 47~K, and the average
production rates are: $3.5\times10^{26}$ molec~s$^{-1}$ for CH$_3$OH
and $1.25\times10^{25}$ molec~s$^{-1}$ for HCN.

\end{enumerate}

The complete material used in this study is available in
Appendix~\ref{Ap-Data}.

\acknowledgments

We thank Geronimo Villanueva for helpful recommendations with regard
to the available molecular catalogs, and Steve Charnley and Karin
\"Oberg for valuable discussions on the formation of cometary ices.
This work was supported by NASA through a Planetary Astronomy
Program grant to D.~J.

{\it Facilities:} \facility{IRAM:30m (EMIR)}.

\appendix

\section{Derivation of Physical Quantities from Spectral Lines}\label{Ap-Model}

The lines were parameterized in terms of their area
\mbox{$\int\!\!T_\mathrm{mB}\mathrm{d}v$} (integrated in the
radial-velocity space) and median velocity $v_0$, and these
parameters were converted into three basic physical quantities:
rotational temperature $T_\mathrm{rot}$, molecular production rate
$Q$, and median radial gas-flow velocity $v_\mathrm{rad}$
(Section~\ref{Sec-Obs}).

It is easy to realize that the median line velocity $v_0$ is close
to the median radial gas-flow velocity within the beam
$v_\mathrm{rad}$, if (\emph{i}) the observed gas is optically thin
and if (\emph{ii}) the \emph{emission coefficient} (possibly
changing across the observed region of the coma) is uncorrelated
with the radial component of gas velocity. For simplicity we
consider the two velocities to be equal, $v_\mathrm{rad} = v_0$.

The line area \mbox{$\int\!\!T_\mathrm{mB}\mathrm{d}v$} can be
converted into the production rate $Q$ using a simple model, which
requires the following additional assumptions: (\emph{iii}) the
energy levels are populated according to the Boltzmann distribution
at a constant temperature $T$, (\emph{iv}) the volume-density of the
molecules is inversely proportional to the square of the
nucleocentric distance, i.e. the photodissociation losses are
negligible, and (\emph{v}) the molecules are isotropically ejected
at a constant rate from a central source and continue to travel at a
constant speed $v_\mathrm{gas}$. Note that assumption (\emph{iii})
implies \emph{thermal equilibrium} in which the rotational
temperature is equal to the kinetic temperature, and therefore we
simply refer to the gas temperature. This assumption also implies
that the \emph{emission coefficient} for a given transition is
constant in the observed region of the coma, which naturally
surpasses assumption (\emph{ii}).

Under these assumptions the production rate $Q$ can be easily
obtained from \mbox{$\int\!\!T_\mathrm{mB}\mathrm{d}v$} using a
simple formula \citep[see][for derivation]{Dra09}:
\begin{equation}\label{Eq1}
Q = \frac{16\pi}{\sqrt{\pi\ln2}} \frac{k}{hc^2}
\frac{\nu_{ul}}{A_{ul}} \frac{Z(T)}{g_u\,e^{-E_u/kT}}
\frac{b\Delta}{D}
v_\mathrm{gas}\,\textstyle\int\!\!T_\mathrm{mB}\mathrm{d}v,
\end{equation}
where $k$ and $h$ are the Boltzmann and Planck constants,
respectively, and $c$ is the speed of light; the molecule has a
temperature-dependent partition function $Z(T) = \sum_i g_i
\exp(-E_i/kT)$; the transition from the upper rotational energy
level $u$ to the lower level $l$ is parameterized by the rest
frequency of the emitted photon $\nu_{ul}$, Einstein coefficient for
spontaneous emission $A_{ul}$, degeneracy of the upper state $g_u$,
and the upper state energy $E_u$; the telescope has a dish diameter
$D$ and a dimensionless parameter $b$ connecting the beam FWHM with
the dish diameter: $\mathrm{FWHM} = b\,c/(D\,\nu_{ul})$; for the
IRAM~30~m telescope we take $D = 30$~m and $b = 1.13$, the latter
derived from the beam sizes given in the online
documentation\footnote{\url{http://www.iram.es/IRAMES/mainWiki/Iram30mEfficiencies}};
finally, $\Delta$ is the topocentric distance.

Equation~(\ref{Eq1}) immediately implies that
\begin{equation}\label{Eq2}
\ln\Bigg(\frac{\nu_{ul}}{g_u
A_{ul}}\textstyle\int\!\!T_\mathrm{mB}\mathrm{d}v\displaystyle\Bigg)
= -\frac{E_u}{k T} + \textstyle const
\end{equation}
for different lines of the same molecule. It provides a convenient
way of determining $T$ from the slope of the linear relation between
$E_u$ and the logarithmic term, and is called the \emph{rotational
diagram} technique \citep[cf.][]{Boc94}.

Our procedure was as follows: (\emph{i}) we first used
Eq.~(\ref{Eq2}) to obtain the temperature $T$ from the areas of the
five lines of CH$_3$OH, then (\emph{ii}) we calculated the partition
functions $Z(T)$ for this temperature, interpolating it linearly in
log--log space from the catalog values available for a range of
temperatures, and finally, (\emph{iii}) we substituted in
Eq.~(\ref{Eq1}) the obtained $T$ and $Z(T)$ to convert the line
areas \mbox{$\int\!\!T_\mathrm{mB}\mathrm{d}v$} into the production
rates~$Q$, assuming $v_\mathrm{gas} = 0.8$~km~s$^{-1}$. The
molecular constants were taken from the sources cited in
Table~\ref{Tab1}, and likewise for the values of the partition
functions. In the case of CH$_3$OH, we converted the areas of the
average line profiles because from Eq.~(\ref{Eq2}) we knew the
theoretical line ratios at a temperature $T$.

While we investigate the behavior of the (rotational) temperature in
time (Fig.~\ref{Fig-EvolTrot}), we calculated the production rates
of both molecules (Fig.~\ref{Fig-EvolProdRates}) using the constant
temperature of 47~K, derived from the mean spectrum of CH$_3$OH
(Figs.~\ref{Fig-AvgMethanol} and~\ref{Fig-AvgRotDiag}). The actual
adopted value is not very important for the presented results
because it simply scales the production rates of a given molecule by
a constant factor and therefore only affects the
molecule-to-molecule ratio. The role of the gas velocity is even
less important as it identically scales the production rates of all
molecules and does not affect the rotational temperature in the
framework of our simple model (Eqs.~\ref{Eq1} and~\ref{Eq2}). This
velocity can be, in principle, retrieved from the line profiles, but
it is not a trivial task for jet-dominated activity \citep{Dra09}.
At this point we refrain from deriving it in the framework of the
above isotropic model (although such an approach has been widely
used in the past), adopting instead a common literature value
$v_\mathrm{gas} = 0.8$~km~s$^{-1}$ suggested for many comets around
the same heliocentric distance and generally consistent with
theoretical predictions. The real issues can be caused by temporal
variations of these quantities, such as the influence of the varying
$T_\mathrm{rot}$ discussed in Section~\ref{Sec-Sources-ProdRates}.

\section{Complete Material Used in This Study}\label{Ap-Data}

\begin{figure}
\epsscale{0.75} \plotone{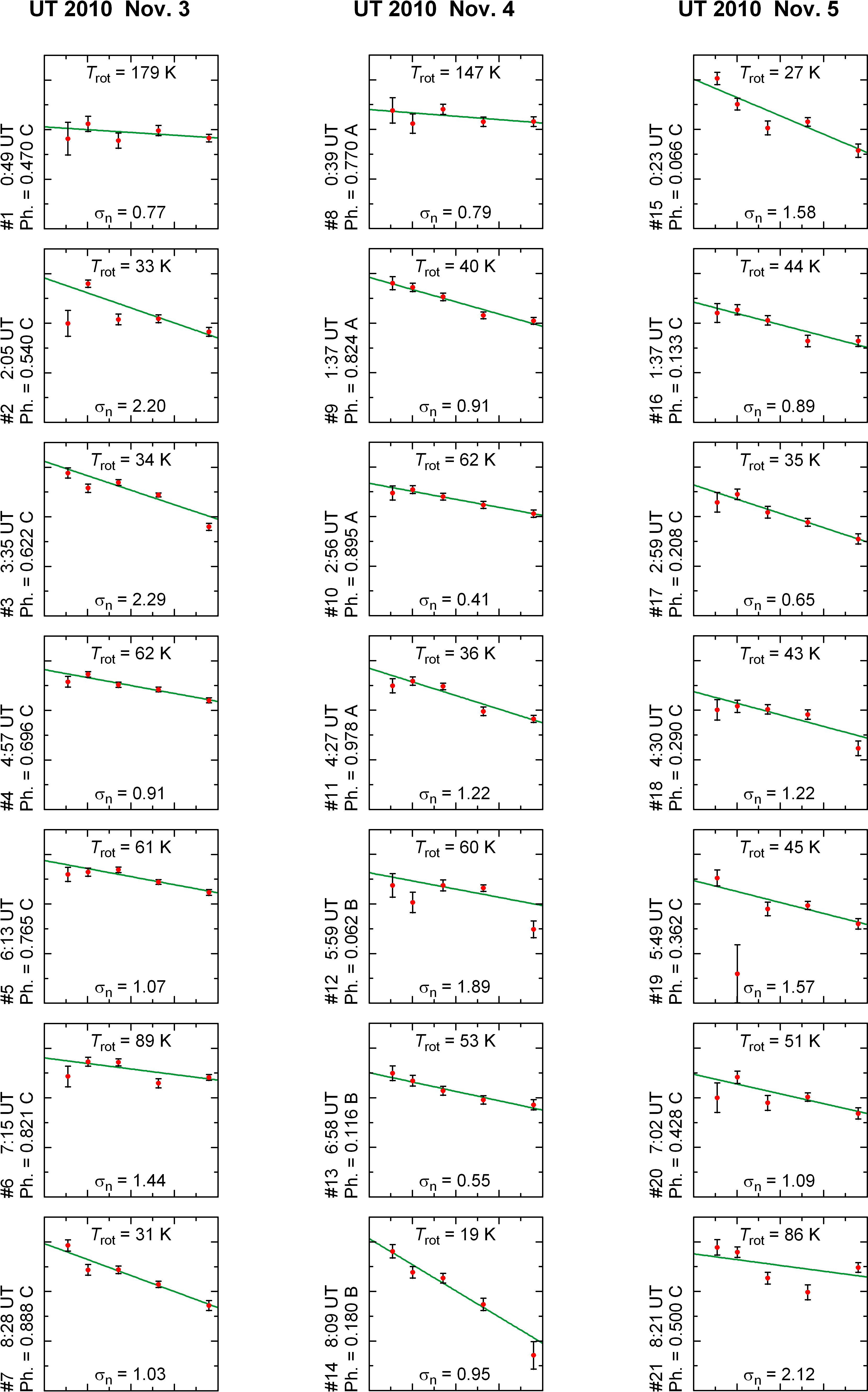} \caption{Full
set of the 21 rotational diagrams derived from the time series of
CH$_3$OH, with weighted linear least-squares fits. Other details,
including the description of the figure axes, are the same as in
Fig.~\ref{Fig-ExRotDiag}. This material is also contained in
Animation~1.\label{Fig-AllRotDiag}}
\end{figure}

\begin{figure}
\epsscale{1.0} \plotone{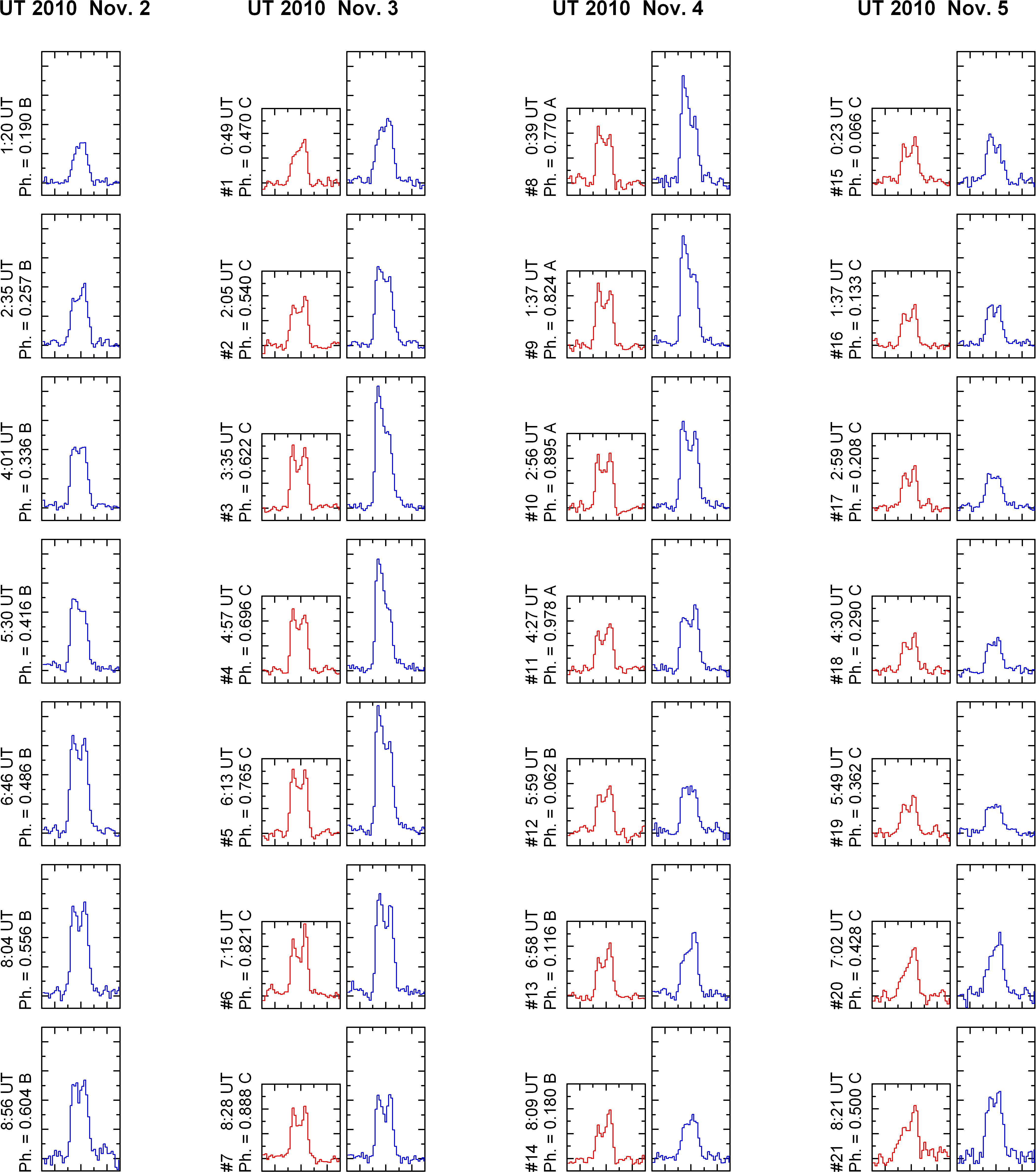} \caption{Full
set of the 28 line profiles of HCN (blue) and 21 line profiles of
CH$_3$OH (red) illustrating the temporal variations. Other details,
including the description of the figure axes, are the same as in
Fig.~\ref{Fig-ExLineShapes}. This material is also contained in
Animation~2.\label{Fig-AllLineShapes}}
\end{figure}

This appendix contains the complete material used in this work. In
Table~\ref{Tab2} we show the time series of all line parameters and
corresponding physical quantities, while Figs.~\ref{Fig-AllRotDiag}
and~\ref{Fig-AllLineShapes} contain all the rotational diagrams and
line profiles, respectively.

\placetable{Tab2}

\placefigure{Fig-AllRotDiag}

\placefigure{Fig-AllLineShapes}

\clearpage

\clearpage

\clearpage

\clearpage

\clearpage

\clearpage

\clearpage

\clearpage

\clearpage

\clearpage

\clearpage

\clearpage

\clearpage

\clearpage

\begin{deluxetable}{llcccccccccc}
\tablewidth{0pt} \tabletypesize{\footnotesize}
\tablecaption{\label{Tab1}Transition and Telescope Constants}
\tablehead{ \colhead{Molecule} & \colhead{Transition} &
\colhead{$\nu_{ul}$\tablenotemark{a}} & $A_{ul}$\tablenotemark{b} &
$g_u$\tablenotemark{c} & $E_u/k$\tablenotemark{d} &
\multicolumn{2}{c}{Beam\tablenotemark{e}} &
\colhead{$\tau_\perp$\tablenotemark{h}} & \colhead{$\Delta
v$\tablenotemark{i}} & \colhead{$\eta_\mathrm{mB}$\tablenotemark{j}}
\\
  & & \colhead{(GHz)} & \colhead{($10^{-6}$~s$^{-1}$)} & & \colhead{(K)} & \colhead{(\arcsec)\tablenotemark{f}} & \colhead{(km)\tablenotemark{g}} & \colhead{(min)} & \colhead{(m~s$^{-1}$)} & }
\startdata
HCN      & \mbox{\textit{J}(3--2)}     & 265.886434 &    835.55 &    21 & 25.521 & 4.4 & 495 & 10.3 & \phn44 & 0.52 \\[1.0ex]
CH$_3$OH & \mbox{$5_{0,3}$--$5_{1,4}$} & 157.178987 & \phn20.38 &    11 & 47.936 & 7.4 & 826 & 17.2 &    149 & 0.72 \\
         & \mbox{$4_{0,3}$--$4_{1,4}$} & 157.246062 & \phn20.98 & \phn9 & 36.335 &     &     &      &        &      \\
         & \mbox{$1_{0,3}$--$1_{1,4}$} & 157.270832 & \phn22.06 & \phn3 & 15.447 &     &     &      &        &      \\
         & \mbox{$3_{0,3}$--$3_{1,4}$} & 157.272338 & \phn21.46 & \phn7 & 27.053 &     &     &      &        &      \\
         & \mbox{$2_{0,3}$--$2_{1,4}$} & 157.276019 & \phn21.82 & \phn5 & 20.090 &     &     &      &        &      \\
\enddata

\tablecomments{Molecular constants for CH$_3$OH are taken from the
\citet{Pea10} update to the JPL Molecular Catalog \citep[][]{Pic98},
available online at \url{http://spec.jpl.nasa.gov}, whereas the
constants for HCN are from the Cologne Database for Molecular
Spectroscopy \citep[][]{Mul05}, available at
\url{http://www.astro.uni-koeln.de/cdms}.}
\tablenotetext{a}{Transition rest frequency.}
\tablenotetext{b}{Einstein coefficient for spontaneous emission,
calculated from the temperature-dependent integrated line intensity
$\mathcal{I}(T)$ using the relation: $A_{ul} = 8\pi
c^{-2}\,\nu_{ul}^2\,Z(T)\,g_u^{-1}\,
[{\exp(-E_l/kT)-\exp(-E_u/kT)}]^{-1}\,\mathcal{I}(T)$ evaluated at
the standard temperature $T=300$~K, where $c$ is the speed of light,
$k$ is the Boltzmann constant, $\nu_{ul}$ is the transition rest
frequency, $Z(T)$ is the temperature-dependent partition function,
$g_u$ is the upper-state degeneracy, and $E_l$ and $E_u$ are the
lower and upper state energies, respectively (note that $E_u - E_l =
h\nu_{ul}$, where $h$ is the Planck constant).}
\tablenotetext{c}{Degeneracy of the upper state.}
\tablenotetext{d}{Upper state energy.} \tablenotetext{e}{Half-width
at half maximum (HWHM) of the beam.} \tablenotetext{f}{Angular beam
size.} \tablenotetext{g}{Beam size at the comet distance.}
\tablenotetext{h}{Minimum escape time from the beam center needed to
reach the HWHM at a constant speed of 0.8~km~s$^{-1}$. Note that
these escape times are much shorter than the photochemical lifetimes
of both molecules \citep[$\sim 20$~hr at this heliocentric distance;
see][]{Hue92}.} \tablenotetext{i}{Native velocity spacing of the
spectral channels.} \tablenotetext{j}{Main-beam efficiency
interpolated from the values given at
\url{http://www.iram.es/IRAMES/mainWiki/Iram30mEfficiencies}.}
\end{deluxetable}

\clearpage

\begin{deluxetable}{cccc|llll|lll}
\tablewidth{0pt} \tabletypesize{\footnotesize} \rotate
\tablecaption{\label{Tab2}Complete Time Series of the Line
Parameters and Corresponding Physical Quantities} \tablehead{
\colhead{\#}&\colhead{UT~Date}&\colhead{$\Delta\mathrm{t}$}&\colhead{Phase}&\multicolumn{4}{c}{CH$_3$OH}&\multicolumn{3}{c}{HCN}\\
\colhead{}&\colhead{(Nov.~2010)}&\colhead{(min)}&\colhead{\&~Cycle}&\colhead{\mbox{$\int\!\!T_\mathrm{mB}\mathrm{d}v$}}&\colhead{$v_0$}&\colhead{$T_\mathrm{rot}$}&\colhead{$\log(Q)$}&\colhead{\mbox{$\int\!\!T_\mathrm{mB}\mathrm{d}v$}}&\colhead{$v_0$}&\colhead{$\log(Q)$}}
\startdata
        & $2.0558$ & $51.2$ & 0.190 B &  \hspace{0.37cm}\nodata   &   \hspace{0.49cm}\nodata   & \hspace{0.14cm}\nodata &   \hspace{0.45cm}\nodata   & $0.672_{-0.021}^{+0.106}$ & $-0.002_{-0.022}^{+0.023}$ & $24.813_{-0.013}^{+0.081}$ \\
        & $2.1075$ & $61.3$ & 0.257 B &  \hspace{0.37cm}\nodata   &   \hspace{0.49cm}\nodata   & \hspace{0.14cm}\nodata &   \hspace{0.45cm}\nodata   & $1.060_{-0.017}^{+0.164}$ & $-0.032_{-0.014}^{+0.014}$ & $25.011_{-0.007}^{+0.078}$ \\
        & $2.1672$ & $61.9$ & 0.336 B &  \hspace{0.37cm}\nodata   &   \hspace{0.49cm}\nodata   & \hspace{0.14cm}\nodata &   \hspace{0.45cm}\nodata   & $1.234_{-0.012}^{+0.190}$ & $-0.117_{-0.012}^{+0.011}$ & $25.076_{-0.004}^{+0.078}$ \\
        & $2.2289$ & $61.6$ & 0.416 B &  \hspace{0.37cm}\nodata   &   \hspace{0.49cm}\nodata   & \hspace{0.14cm}\nodata &   \hspace{0.45cm}\nodata   & $1.357_{-0.013}^{+0.219}$ & $-0.180_{-0.013}^{+0.012}$ & $25.118_{-0.004}^{+0.082}$ \\
        & $2.2817$ & $61.0$ & 0.486 B &  \hspace{0.37cm}\nodata   &   \hspace{0.49cm}\nodata   & \hspace{0.14cm}\nodata &   \hspace{0.45cm}\nodata   & $1.890_{-0.020}^{+0.290}$ & $-0.120_{-0.011}^{+0.011}$ & $25.262_{-0.004}^{+0.077}$ \\
        & $2.3359$ & $60.4$ & 0.556 B &  \hspace{0.37cm}\nodata   &   \hspace{0.49cm}\nodata   & \hspace{0.14cm}\nodata &   \hspace{0.45cm}\nodata   & $1.786_{-0.031}^{+0.285}$ & $-0.068_{-0.020}^{+0.022}$ & $25.237_{-0.007}^{+0.081}$ \\
        & $2.3723$ & $20.7$ & 0.604 B &  \hspace{0.37cm}\nodata   &   \hspace{0.49cm}\nodata   & \hspace{0.14cm}\nodata &   \hspace{0.45cm}\nodata   & $1.509_{-0.041}^{+0.257}$ & $-0.049_{-0.029}^{+0.032}$ & $25.164_{-0.012}^{+0.087}$ \\[1.0ex]
\phn$1$ & $3.0339$ & $61.1$ & 0.470 C & $0.070_{-0.003}^{+0.007}$ & $-0.043_{-0.034}^{+0.037}$ &  $179_{-86}^{+816}$    & $26.440_{-0.020}^{+0.044}$ & $1.222_{-0.024}^{+0.186}$ & $-0.033_{-0.014}^{+0.013}$ & $25.072_{-0.008}^{+0.077}$ \\
\phn$2$ & $3.0868$ & $61.9$ & 0.540 C & $0.088_{-0.003}^{+0.007}$ & $-0.091_{-0.035}^{+0.035}$ &  \phn$33_{-4}^{+5}$    & $26.534_{-0.017}^{+0.040}$ & $1.503_{-0.016}^{+0.232}$ & $-0.152_{-0.011}^{+0.011}$ & $25.162_{-0.005}^{+0.079}$ \\
\phn$3$ & $3.1495$ & $61.5$ & 0.622 C & $0.118_{-0.002}^{+0.009}$ & $-0.114_{-0.026}^{+0.026}$ &  \phn$34_{-3}^{+3}$    & $26.664_{-0.009}^{+0.035}$ & $1.916_{-0.017}^{+0.296}$ & $-0.249_{-0.008}^{+0.008}$ & $25.268_{-0.004}^{+0.078}$ \\
\phn$4$ & $3.2062$ & $61.7$ & 0.696 C & $0.115_{-0.003}^{+0.009}$ & $-0.135_{-0.024}^{+0.022}$ &  \phn$62_{-8}^{+11}$   & $26.653_{-0.011}^{+0.036}$ & $1.720_{-0.014}^{+0.282}$ & $-0.263_{-0.007}^{+0.008}$ & $25.221_{-0.004}^{+0.084}$ \\
\phn$5$ & $3.2593$ & $61.7$ & 0.765 C & $0.123_{-0.003}^{+0.010}$ & $-0.168_{-0.027}^{+0.027}$ &  \phn$61_{-9}^{+11}$   & $26.681_{-0.012}^{+0.038}$ & $2.154_{-0.019}^{+0.343}$ & $-0.171_{-0.008}^{+0.009}$ & $25.319_{-0.004}^{+0.083}$ \\
\phn$6$ & $3.3020$ & $30.2$ & 0.821 C & $0.124_{-0.004}^{+0.010}$ & $-0.037_{-0.046}^{+0.044}$ &  \phn$89_{-22}^{+32}$  & $26.685_{-0.014}^{+0.037}$ & $1.869_{-0.021}^{+0.293}$ & $-0.085_{-0.013}^{+0.013}$ & $25.257_{-0.005}^{+0.080}$ \\
\phn$7$ & $3.3527$ & $74.9$ & 0.888 C & $0.097_{-0.003}^{+0.008}$ & $-0.152_{-0.036}^{+0.039}$ &  \phn$31_{-3}^{+4}$    & $26.578_{-0.015}^{+0.037}$ & $1.115_{-0.013}^{+0.180}$ & $-0.124_{-0.022}^{+0.021}$ & $25.032_{-0.005}^{+0.082}$ \\[1.0ex]
\phn$8$ & $4.0268$ & $30.0$ & 0.770 A & $0.093_{-0.004}^{+0.009}$ & $-0.224_{-0.042}^{+0.037}$ &  $147_{-69}^{+385}$    & $26.560_{-0.020}^{+0.045}$ & $1.595_{-0.028}^{+0.243}$ & $-0.235_{-0.014}^{+0.014}$ & $25.188_{-0.007}^{+0.077}$ \\
\phn$9$ & $4.0676$ & $60.5$ & 0.824 A & $0.110_{-0.004}^{+0.008}$ & $-0.176_{-0.034}^{+0.035}$ &  \phn$40_{-5}^{+6}$    & $26.633_{-0.015}^{+0.036}$ & $1.775_{-0.019}^{+0.271}$ & $-0.215_{-0.009}^{+0.010}$ & $25.234_{-0.005}^{+0.077}$ \\
   $10$ & $4.1222$ & $61.3$ & 0.895 A & $0.098_{-0.003}^{+0.007}$ & $-0.165_{-0.029}^{+0.029}$ &  \phn$62_{-11}^{+14}$  & $26.584_{-0.013}^{+0.035}$ & $1.577_{-0.017}^{+0.250}$ & $-0.124_{-0.012}^{+0.011}$ & $25.183_{-0.005}^{+0.081}$ \\
   $11$ & $4.1852$ & $61.3$ & 0.978 A & $0.093_{-0.003}^{+0.008}$ & $-0.059_{-0.031}^{+0.038}$ &  \phn$36_{-4}^{+5}$    & $26.559_{-0.014}^{+0.039}$ & $1.186_{-0.015}^{+0.191}$ & $-0.026_{-0.016}^{+0.018}$ & $25.059_{-0.005}^{+0.083}$ \\
   $12$ & $4.2492$ & $30.3$ & 0.062 B & $0.089_{-0.004}^{+0.008}$ & $-0.114_{-0.044}^{+0.048}$ &  \phn$60_{-13}^{+20}$  & $26.541_{-0.019}^{+0.041}$ & $0.889_{-0.021}^{+0.146}$ & $-0.082_{-0.024}^{+0.025}$ & $24.934_{-0.010}^{+0.085}$ \\
   $13$ & $4.2904$ & $60.8$ & 0.116 B & $0.081_{-0.003}^{+0.007}$ & $+0.017_{-0.035}^{+0.034}$ &  \phn$53_{-11}^{+16}$  & $26.500_{-0.016}^{+0.043}$ & $0.978_{-0.020}^{+0.160}$ & $+0.020_{-0.020}^{+0.020}$ & $24.976_{-0.009}^{+0.085}$ \\
   $14$ & $4.3399$ & $60.4$ & 0.180 B & $0.069_{-0.004}^{+0.006}$ & $-0.007_{-0.043}^{+0.044}$ &  \phn$19_{-2}^{+3}$    & $26.432_{-0.023}^{+0.043}$ & $0.753_{-0.019}^{+0.125}$ & $-0.004_{-0.024}^{+0.023}$ & $24.862_{-0.011}^{+0.087}$ \\[1.0ex]
   $15$ & $5.0163$ & $60.5$ & 0.066 C & $0.090_{-0.004}^{+0.008}$ & $+0.006_{-0.052}^{+0.056}$ &  \phn$27_{-3}^{+5}$    & $26.545_{-0.021}^{+0.043}$ & $0.795_{-0.024}^{+0.135}$ & $-0.225_{-0.033}^{+0.026}$ & $24.886_{-0.013}^{+0.088}$ \\
   $16$ & $5.0673$ & $60.3$ & 0.133 C & $0.067_{-0.003}^{+0.006}$ & $-0.078_{-0.039}^{+0.044}$ &  \phn$44_{-8}^{+12}$   & $26.420_{-0.020}^{+0.041}$ & $0.762_{-0.019}^{+0.117}$ & $-0.123_{-0.029}^{+0.026}$ & $24.867_{-0.011}^{+0.078}$ \\
   $17$ & $5.1246$ & $61.4$ & 0.208 C & $0.075_{-0.003}^{+0.006}$ & $-0.057_{-0.051}^{+0.055}$ &  \phn$35_{-5}^{+7}$    & $26.467_{-0.019}^{+0.040}$ & $0.670_{-0.013}^{+0.101}$ & $-0.147_{-0.023}^{+0.025}$ & $24.811_{-0.009}^{+0.076}$ \\
   $18$ & $5.1873$ & $61.3$ & 0.290 C & $0.061_{-0.003}^{+0.006}$ & $-0.065_{-0.050}^{+0.051}$ &  \phn$43_{-7}^{+9}$    & $26.375_{-0.021}^{+0.046}$ & $0.521_{-0.016}^{+0.084}$ & $-0.044_{-0.028}^{+0.030}$ & $24.702_{-0.013}^{+0.082}$ \\
   $19$ & $5.2426$ & $61.3$ & 0.362 C & $0.063_{-0.003}^{+0.006}$ & $-0.076_{-0.044}^{+0.060}$ &  \phn$45_{-10}^{+17}$  & $26.394_{-0.022}^{+0.045}$ & $0.578_{-0.021}^{+0.092}$ & $-0.083_{-0.033}^{+0.034}$ & $24.747_{-0.015}^{+0.081}$ \\
   $20$ & $5.2932$ & $60.8$ & 0.428 C & $0.075_{-0.004}^{+0.007}$ & $+0.004_{-0.034}^{+0.033}$ &  \phn$51_{-11}^{+17}$  & $26.467_{-0.021}^{+0.043}$ & $1.009_{-0.022}^{+0.165}$ & $+0.028_{-0.024}^{+0.025}$ & $24.989_{-0.009}^{+0.084}$ \\
   $21$ & $5.3477$ & $60.3$ & 0.500 C & $0.107_{-0.004}^{+0.010}$ & $+0.074_{-0.040}^{+0.049}$ &  \phn$86_{-29}^{+62}$  & $26.623_{-0.016}^{+0.044}$ & $1.216_{-0.044}^{+0.197}$ & $-0.042_{-0.031}^{+0.034}$ & $25.070_{-0.015}^{+0.083}$
\enddata
\tablecomments{$\Delta\mathrm{t}$ is the time coverage of the
measurement; other symbols and units are the same as defined and
used throughout the paper. The errors on
\mbox{$\int\!\!T_\mathrm{mB}\mathrm{d}v$} and on the resulting $Q$
include the uncertainty of the telescope pointing, and therefore are
partly correlated in the simultaneous measurements of CH$_3$OH and
HCN, unlike the errors on subsequent data points. The errors on
$v_0$ and $T_\mathrm{rot}$ are free of the pointing contribution and
therefore are fully independent, both in time and between the
molecules.}
\end{deluxetable}

\end{document}